\documentclass[reprint,twocolumn,secnumarabic,amssymb, aps, prl]{revtex4-1}

\usepackage{graphicx}
\usepackage{gensymb}
\usepackage{amsmath}
\usepackage[colorlinks,linkcolor=black,citecolor=black,urlcolor=black,filecolor=black]{hyperref}
\usepackage{wrapfig}
\begin{document}

\title{Identifying and mitigating charge instabilities in shallow diamond nitrogen-vacancy centers}

\author{Dolev Bluvstein, Zhiran Zhang, and Ania C. Bleszynski Jayich}
\affiliation{Department of Physics, University of California, Santa Barbara, CA 93106, USA}

\begin{abstract}
The charge degree of freedom in solid-state defects fundamentally underpins the electronic spin degree of freedom, a workhorse of quantum technologies. Here we study charge state properties of individual near-surface nitrogen-vacancy (NV) centers in diamond, where NV$^{-}$ hosts the metrologically relevant electron spin. We find that NV$^{-}$ initialization fidelity varies between individual centers and over time, and we alleviate the deleterious effects of reduced NV$^{-}$ initialization fidelity via logic-based initialization. We also find that NV$^{-}$ can ionize in the dark, which compromises spin measurements but is mitigated by measurement protocols we present here. We identify tunneling to a single, local electron trap as the mechanism for ionization in the dark and we develop NV-assisted techniques to control and readout the trap charge state. Our understanding and command of the NV's local electrostatic environment will simultaneously guide materials design and provide novel functionalities with NV centers.
\end{abstract}

\maketitle
Solid-state defects are important tools in quantum technologies. Prominent examples include nitrogen-vacancy \cite{Doherty2013,Childress2006,Jelezko2004,Gruber1997} and silicon-vacancy \cite{Rose2018,Muller2014} centers in diamond as well as various defects in silicon carbide \cite{Christle2015,Koehl2011,Widmann2015}, where the electronic spin degree of freedom is commonly employed for quantum tasks such as sensing or computing. Importantly, these defects also harbor a charge degree of freedom. The charge degree of freedom sets the number of unpaired electrons that constitute the spin degree of freedom, and so control over spin necessitates control over charge. Lack of charge control can lead to deleterious effects on the defect's functionality as a qubit or sensor. However, with sufficient understanding and control, the charge degree of freedom can also be harnessed for a variety of applications such as high fidelity spin readout \cite{Shields2015,Wolfowicz2017,Hopper2016,Siyushev2013}, super-resolution microscopy \cite{Gu2013,Bersin2018a,Chen2015b}, enhancing quantum coherence \cite{Rose2018,Pfender2017}, and electrical sensing modalities \cite{Karaveli2016,Wolfowicz2018}. 

Shallow, negatively charged nitrogen-vacancy (NV$^-$) centers in diamond have received particular attention for their sensing prowess, recently demonstrating nanoscale magnetic imaging of condensed matter \cite{Thiel2016,Pelliccione2016a} and biological systems \cite{Shi2015,Staudacher2013,Lovchinsky2016}, thermal imaging \cite{Kucsko2013a,Neumann2013}, and electrical conductivity imaging \cite{Ariyaratne2018a} at the nanoscale. Shallow NV$^-$ centers can also interface with other quantum elements in hybrid quantum systems \cite{Faraon2011,Lee2017,Cai2013}. On the other hand, neutral NV$^0$ centers have not achieved promising electron spin control but are commonly observed \cite{Gaebel2006,Iakoubovskii2000,Manson2005a} and result in undesired background in NV$^-$ experiments. Of particular importance for shallow NV$^-$ centers is that the diamond surface is observed to preferentially convert NV$^{-}$ to NV$^{0}$ \cite{Hauf2011,Rondin2010,Fu2010}, thus imposing a clear obstacle to nanoscale sensing applications, where the NV depth is critical to both sensitivity and spatial resolution \cite{Taylor2008a,Rugar2015}.

\begin{figure}
\includegraphics[width=86mm]{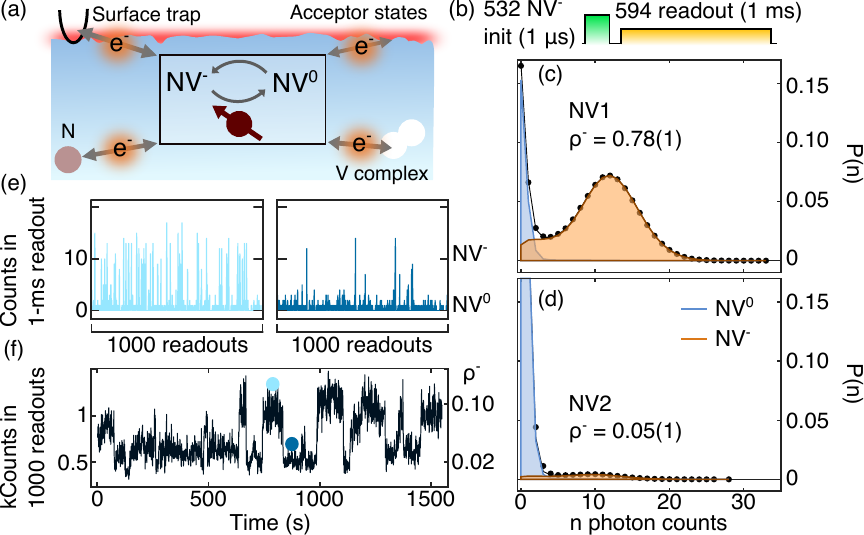}
\caption{NV charge state characteristics vary with different local charge environments. (a) The charge state of local nitrogen (N) centers, vacancy (V) complexes, surface electron traps, and surface acceptor states can all affect the NV charge state. (b) Pulse sequence to measure NV$^{-}$ initialization fidelity with 532-nm illumination. (c),(d) Probability of measuring $n$ photons P$(n)$ for charge stable (c) and charge unstable (d) NV centers in the same sample. Black curve is the sum of the fitted NV$^-$ and NV$^0$ distributions. $\rho^{-}$ is the probability to be in NV$^{-}$ immediately after the 532-nm initialization in (b). (e) Two sets of 1000 consecutive 1-ms-long measurements from the same data comprising the distribution in (d). (f) Consecutive measurements binned into sets of 1000 (1 second each).
}
\label{fig1}
\end{figure}

Under optical illumination in bulk diamond, single NV centers continuously interconvert between negative and neutral charge states as the NV exchanges electrons with the electronic bands, where the steady-state NV$^{-}$ population reaches $\approx$ 75\% under commonly used CW 532-nm excitation \cite{Aslam2013,Shields2015,Chen2013,Doi2016,Doi2014,Hopper2016}. For near-surface NV centers, however, understanding of photoinduced charge interconversion is largely limited to ensemble measurements which explain surface-induced NV$^{-}$ ionization as a result of upwards band bending from surface acceptor states \cite{Hauf2011,Newell2016}. In the dark, recent studies on NV ensembles have shown that NV charge states can be both stable \cite{Dhomkar2018,Dhomkar2016,Jayakumar2016} and unstable \cite{Giri2018,Choi2017,Dhomkar2018a}. Instability of shallow NV$^-$ centers under illumination or in the dark can directly compromise computing and sensing modalities, yet understanding is still limited.

In this work we study the charge state properties of single, shallow NV centers both under illumination and in the dark, focusing in particular on the implications for sensing and on identifying the microscopic origins of charge state instability. We find that the fidelity of optical initialization into NV$^{-}$ exhibits large variations between shallow NV centers as well as over time. We identify reduced NV$^{-}$ initialization fidelity as the primary reduction to spin measurement contrast in shallow NVs, which we alleviate by implementing logic-based charge initialization. We also find that shallow NV$^{-}$ centers can ionize to NV$^{0}$ in the dark, which we methodically identify as tunneling to a single, local electron trap. We achieve control and readout of the trap charge state and measure its optical ionization properties. Further, we show that charge conversion in the dark can produce anomalous signatures in spin measurements and, at worst, will appear indistinguishable from $T_1$ and $T_2$ spin decay; we relieve this detrimental effect by measurement protocols we present here.

The experimental setup consists of a homebuilt, room-temperature confocal microscope for optically addressing NV$^-$ and NV$^0$ centers, which have zero-phonon lines at 637 nm and 575 nm, respectively \cite{Iakoubovskii2000}. We use a 532-nm laser for spin and charge state initialization and for NV$^-$ spin state readout, and we use a 594-nm laser for charge state readout. Under $\sim 2$ $\mu$W of 594-nm excitation, NV$^{-}$ is $\sim$ 40x brighter than NV$^{0}$ in our setup. NV centers are formed by $^{14}$N ion implantation at 4 keV with a dosage of $5.2\times 10^{10}$  ions / cm$^2$ into a 150-$\mu$m thick Element 6 electronic grade (100) diamond substrate, followed by subsequent annealing at 850$\degree$ C for 2.5 hours (see Supplementary Information SI Note 1 \footnote{See the Supplemental Material.} for full details on sample preparation). The NV center depth is experimentally measured via proton NMR \cite{Pham2016,Ajoy2017} and ranges between $\sim$ 3-17 nm (see SI Fig.~S5 \cite{Note1}).

We first report on NV$^{-}$ initialization fidelity $\rho^-$ under 532-nm excitation and its variation in near-surface NV centers. $\rho^-$ is an important parameter for it directly affects NV$^{-}$ measurement sensitivity; the NV$^0$ state gives unwanted background while not contributing to the sensing signal. Here we find that $\rho^-$ varies strongly for shallow NV centers and can be significantly less than 75\%, the commonly reported value for bulk NV centers \cite{Aslam2013,Shields2015,Chen2013,Doi2016,Doi2014,Hopper2016}. In Figs.~\ref{fig1}(c) and \ref{fig1}(d) we measure the NV$^{-}$ initialization fidelity $\rho^{-}$ for two near-surface NVs in the same sample. Plotted are the statistics for the number of photons measured during a 1-ms-long 594-nm readout pulse following a 532-nm initialization pulse. The photon statistics are fit to the model in SI Note 2.1 \cite{Note1}, which is approximately the sum of two Poisson distributions for NV$^{-}$ and NV$^{0}$ \cite{Shields2015,Hacquebard2018}. The relative contribution of the NV$^{-}$ distribution yields $\rho^{-}$. For NV1 presented in Fig.~\ref{fig1}(c), we extract $\rho^{-}$ = 0.78(1), reproducing the typical reported value for bulk NVs. In contrast, we measure $\rho^{-}$ = 0.05(1) for NV2, shown in Fig.~\ref{fig1}(d). From a sample of 67 individual centers we measure $\langle \rho^- \rangle = 0.59$ and $\sigma_{\rho^-} = 0.15$ (see SI Fig.~S1 \cite{Note1}).

We also find that $\rho^{-}$ can vary in time on timescales spanning seconds to months. To capture the faster dynamics, in Fig.~\ref{fig1}(e) we plot two data sets, each consisting of one thousand consecutive 1-ms-long readouts on NV2. The two data sets, taken 2 seconds apart, show a notable difference in outcomes of NV$^-$, as measured by photon counts, indicating that $\rho^-$ is larger in the first data set than in the second. Coarse-graining the data by binning one thousand consecutive measurements yields the data in Fig.~\ref{fig1}(f), which shows that $\rho^{-}$ takes on discrete values that are stable on timescales of seconds to minutes. This discrete behavior suggests that the NV charge state is governed by  discrete, metastable configurations of the local charge environment. 
In practice, this environment-induced slow blinking, which is also observed under CW 532-nm excitation and is distinct from photoinduced hopping between NV charge states, can reduce the sensitivity of near-surface NV centers by introducing substantial slow noise into measurements.

\begin{figure}
\includegraphics[width=86mm]{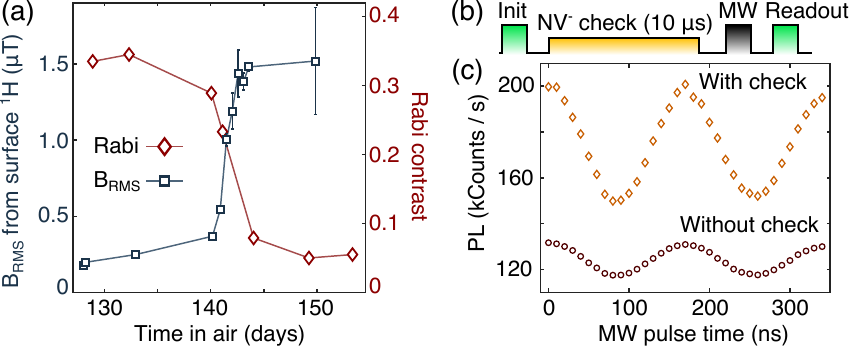}
\caption{(a) Rabi contrast and root-mean-square magnetic field $B_{\text{RMS}}$ produced by surface $^1$H as a function of time under ambient conditions.  Rabi contrast is defined as the peak-to-peak amplitude of the PL oscillations divided by the maximum PL. (b) Rabi pulse sequence for (c) including logic-based charge initialization protocol. (c) Photoluminescence-based-measurement of Rabi oscillations on charge unstable NV center ($\rho^{-} \approx 0.15$), with (orange diamonds) and without (red circles) using precheck protocol.}
\label{fig2}
\end{figure}

Our measurements also reveal that the average $\rho^{-}$ decreases on the time scale of days to months, and this decrease is strongly correlated to environmental changes at the diamond surface. As a practical metric for $\rho^-$ we monitor the NV$^-$ spin fluorescence contrast in a Rabi oscillation measurement; the contrast is reduced when the NV spends more time in the neutral NV$^0$ state, which contributes spin-independent background fluorescence. Fig.~\ref{fig2}(a) plots the Rabi contrast of NV1 as a function of time after a standard surface preparation protocol consisting of acid cleaning and oxygen annealing (see SI Note 1 \cite{Note1}). The Rabi contrast was stable at $35\%$ for 130 days before suddenly decreasing to 5\% over a span of 20 days. Other NV centers exhibit similar behavior, with \textit{e.g.} NV2 exhibiting a drop in $\rho^{-}$ from $\approx 75\%$ to $5\%$ over several months. Notably, cleaning the surface induces a partial or full recovery of $\rho^-$, suggesting that changes in $\rho^-$ are dominated by surface effects. 

Critically, as shown in Fig.~\ref{fig2}(a), we find that the reduction in Rabi contrast on NV1 (measured depth $\sim$ 3.5 nm) is strongly correlated with an increase in the number of $^1$H nuclear spins on the diamond surface, as measured via NV-based nuclear magnetic resonance \cite{Staudacher2013,Mamin2013}. The root-mean-square magnetic field $B_{\text{RMS}}$ produced by surface $^1$H is measured with an XY8-k sensing sequence (see SI Fig.~S5 \cite{Note1}) \cite{Pham2016,Loretz2015}. The reason for the increased hydrogen is unclear, but we make a few observations. The 1.5 $\mu T$ $B_{\text{RMS}}$ value measured after long air exposure is too large to be exclusively due to a two-dimensional surface hydrogen termination layer, indicating that other adsorbates such as water or hydrocarbons are contributing. Further, other NV centers did not exhibit similar changes in contrast and $^1$H density between days 130 - 150, and hence we speculate that laser illumination plays a role as we illuminated only NV1 during that period. Maintaining high Rabi contrast over extended periods of time is critical for NV-based sensing, and the correlation discovered here motivates a more detailed investigation.

The deleterious effects of low $\rho^-$ on Rabi contrast can be alleviated by implementing a measurement protocol [Fig.~\ref{fig2}(b)] that checks for successful NV$^-$ initialization prior to the spin measurement sequence. In Fig.~\ref{fig2}(c) we plot a Rabi measurement with and without this precheck; the spin measurement contrast increases from 14 kCounts/s to 50 kCounts/s and the measured signal-to-noise ratio increases 3-fold. This result also serves to confirm that poor NV$^-$ initialization fidelity is the dominant source of reduced Rabi contrast. In demonstrating this precheck technique in Fig.~\ref{fig2}(b), we postselect on the raw data by removing measurements where no photons are detected during the 10-$\mu s$, 594-nm NV$^{-}$ check in Fig.~\ref{fig2}(b). In practice, to increase measurement sensitivity one would integrate on-the-fly logic to reinitialize after a failed precheck.

\begin{figure}
\includegraphics[width=86mm]{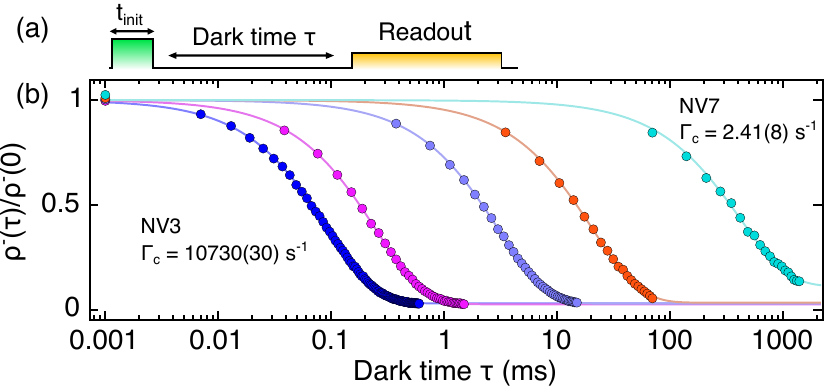}
\caption{NV$^{-}$ survival probability in the dark after 532-nm initialization. (a) Measurement sequence for (b). (b) NV$^{-}$ ionization in the dark measured on 5 representative NV centers; solid curves are fits to the exponential decay in Eq.~(\ref{fitmodel}). Left to right, $\Gamma_c = 10730(30)$, $4030(10)$, $331(1)$, $48.9(6)$, $2.41(8) ~\text{s}^{-1}$.}
\label{fig3}
\end{figure}

We now turn to a discussion of NV charge state dynamics in the dark. In Fig.~\ref{fig3}(b) we plot the NV$^{-}$ population as a function of dark wait time after a 532-nm initialization pulse [Fig.~\ref{fig3}(a)] for five NVs; we find NV$^{-}$ ionizes to NV$^{0}$ in the dark with a wide distribution of decay times. All NVs fit well to a model of exponential decay

\begin{flalign}
\rho^{-}(t) / \rho^{-}(0) = 1 - A \left(1-e^{-\Gamma_c t} \right)
\label{fitmodel}
\end{flalign}

where decay rate $\Gamma_c$, starting NV$^{-}$ population $\rho^{-}(0)$, and decay amplitude $A$ are free fit parameters. The five NVs plotted in Fig.~\ref{fig3} span four orders of magnitude in fitted $\Gamma_c$, with timescales ranging from 100 $\mu$s to seconds. From a sample of 108 individual centers, approximately: 10\% of NVs have $\Gamma_c > 50 ~\text{s}^{-1}$, 10\% have $50 ~\text{s}^{-1}> \Gamma_c > 20~\text{s}^{-1}$, 30\% have $20~\text{s}^{-1} > \Gamma_c > 1~\text{s}^{-1}$, and 50\% have $\Gamma_c < 1~\text{s}^{-1}$. We do not observe a dependence of $\Gamma_c$ on magnetic field or a strong correlation with NV depth (see SI Table S1 \cite{Note1}). See SI Note 2.2 \cite{Note1} for details of measuring $\rho^{-}(t)$.

\begin{figure}
\includegraphics[width=86mm]{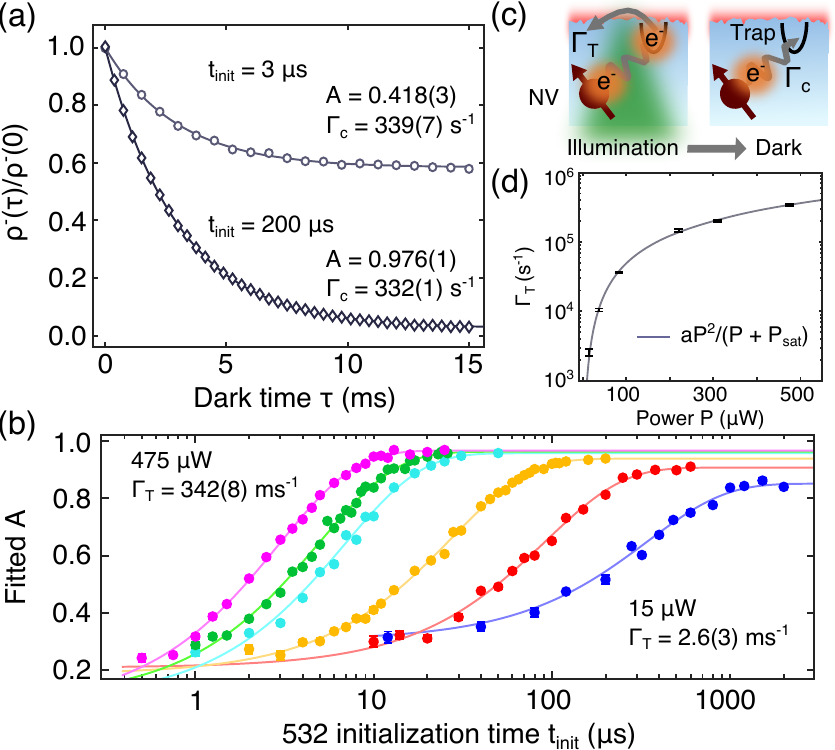}
\caption{(a) Charge decay measurement [Fig.~\ref{fig3}(a)] with 225-$\mu$W, 532-nm excitation for t$_{\text{init}}$ = 3 $\mu$s (light circles) and 200 $\mu$s (dark diamonds). Solid curves are fits to the model in Eq.~(\ref{fitmodel}). (b) Fitted values of $A$ for varied initialization times at six laser powers ranging from 15 $\mu$W to 475 $\mu$W. The solid curves are fits to a model of exponential saturation with rate $\Gamma_T + \Gamma_c$. (c) Model for charge decay: a local electron trap is ionized during initialization at rate $\Gamma_T$ and captures an electron from NV$^{-}$ at fixed tunneling rate $\Gamma_c$. $A$ is then the probability that the trap is empty. (d) Trap ionization rate $\Gamma_T$ vs. laser power. Solid curve is a fit to $aP^2 / (P + P_{\text{sat}})$.}
\label{fig4}
\end{figure}

We find the dark ionization process is highly dependent on initialization power and duration. Figure~\ref{fig4}(a) plots the charge decay observed on NV5 [middle curve in Fig.~\ref{fig3}(b)] for different initialization times t$_{\text{init}} = 3 ~\text{and} ~200 ~\mu$s. Interestingly, the two fits yield the same value of $\Gamma_c$ (within error), but $A$ changes substantially; as $t \rightarrow \infty$, NV$^{-}$ decays to NV$^{0}$ in 98\% of the measurement shots for t$_{\text{init}}$ = 200 $\mu$s, but only in 42\% of the shots with t$_{\text{init}}$ = 3 $\mu$s. To arrive at a more quantitative understanding we repeat the measurement in Fig.~\ref{fig4}(a), varying $t_{\text{init}}$ over a large range of values. The dependence of $A$ on $t_{\text{init}}$ is plotted in Fig.~\ref{fig4}(b) at six laser powers, and the result is fit well by an exponential with a rate that increases with power. $\Gamma_c$ does not change with t$_{\text{init}}$ or power (see SI Fig.~S2 \cite{Note1}). We note that in Fig.~\ref{fig3}, the laser power and t$_{\text{init}}$ were chosen on each NV such that $A$ reaches its saturation value. 

To explain the observations of Figs.~\ref{fig4}(a) and \ref{fig4}(b), in Fig.~\ref{fig4}(c) we present a model of electron tunneling to a single local electron trap with fixed tunneling rate $\Gamma_c$. If the trap is empty, NV$^{-}$ will decay to NV$^{0}$ as $t \rightarrow \infty$; $A$ then represents the probability that the trap is empty. In our model, the green illumination ionizes the trap at rate $\Gamma_T$ and thus empties the trap with probability $A \sim 1 -\exp{\left(-\Gamma_T \text{t$_{\text{init}}$}\right)}$, as observed in Fig.~\ref{fig4}(b). To repump the trap between measurement repetitions, we optically initialize into NV$^-$ and wait in the dark for a time $> 3 / \Gamma_c$. A key result is that the presence of multiple dominant traps is inconsistent with the data in Fig.~\ref{fig4}. More than one dominant trap would result in a non-mono-exponential decay and necessitates that the fitted $\Gamma_c$ increase with $A$, which we do not observe (see SI Note 3 \cite{Note1}). With this analysis we identify the mechanism for charge decay as tunneling to a single local electron trap. Moreover, we can quantitatively set the trap charge state population by varying $t_{\text{init}}$ as in Fig.~\ref{fig4}(b).

We probe the ionization properties of the trap with our NV-assisted control and readout capabilities: we intentionally populate the trap via tunneling from NV$^{-}$ to trap, then ionize the trap optically while repopulating NV$^-$, and finally measure the trap charge state via NV$^{-}$ ionization in the dark. In Fig.~\ref{fig4}(d) we plot the trap ionization rate $\Gamma_T$ versus 532-nm laser power. We find $\Gamma_T$ is fit well by a saturation model of $a P^2 / (P + P_{\text{sat}})$, where $a$ = 814(43) s$^{-1} / \mu\text{W}$, $P$ is laser power, and the saturation power $P_{\text{sat}} = 65(23)~ \mu$W. This power-dependence is consistent with trap ionization by a two-photon transition through an orbital excited state; we note that NV$^-$ requires the energy of two 532-nm photons to photoionize \cite{Aslam2013}, and we expect the trap is lower in energy than NV$^-$. In Fig.~\ref{fig4}(b) we also observe that $A(t_{\text{init}} = \infty)$ increases with laser power, which is qualitatively reproduced by $\Gamma_T / (\Gamma_T + \Gamma_c)$ as a consequence of the rate equations under illumination (see SI Fig.~S6 \cite{Note1}). Physical trap candidates where tunneling could be energetically favorable include vacancy-related complexes, specifically divacancy \cite{Dhomkar2018,Deak2014} and surface sp$^2$ defects \cite{Stacey2018}.

\begin{figure}
\includegraphics[width=86mm]{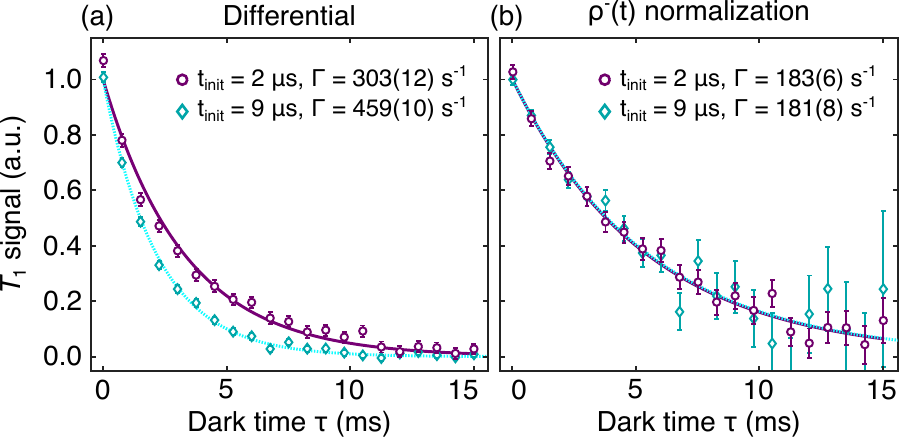}
\caption{Removing the effect of NV$^-$ ionization in the dark from spin measurements. 
(a) Differential $T_1$ measurement, measured with a spin-to-charge sequence \cite{Shields2015,Ariyaratne2018a}. Initialization times of $2 ~\mu$s (purple circles) and $9 ~\mu$s (blue diamonds) yield inconsistent signals. Curves are fits to $\exp(-\Gamma t)$. (b) Normalizing for charge decay during the dark $\tau$ time for the signals in (a) renders agreement between the two measurements and yields $\Gamma$ consistent with spin relaxation rates of other NVs in the same sample ($\Gamma_1 \sim 200 \pm 30 ~\text{s}^{-1}$).}
\label{fig5}
\end{figure}

We now turn to a discussion of the detrimental effects of charge conversion in the dark on spin measurements, as well as the appropriate mitigation protocols. In Fig.~\ref{fig5}(a) an exponential fit to a typical $T_1$ measurement on NV5 yields different relaxation rates depending on the duration of the green initialization pulse, indicating the presence of confounding effects that mask the real $T_1$. The $T_1$ measurement in Fig.~\ref{fig5}(a) employs a common-mode rejection technique referred to as a differential measurement \cite{Myers2017a}, which alleviates effects of recombination in the dark, but importantly does not alleviate ionization in the dark (see SI Note 4 \cite{Note1}). A differential $T_1$ measurement in the case of ionization in the dark, as in Fig.~\ref{fig5}(a), yields

\begin{flalign}
\text{PL$_{\text{diff},T_1}$}(t) = C \exp{\left(-\Gamma_1 t \right)} \left(\rho^{-}(t) / \rho^{-}(0)\right)
\label{diffT1signal}
\end{flalign}

where $\Gamma_1 \equiv 1 / T_1$ and $C$ describes the contrast between the spin states. For NV5 presented in Fig.~\ref{fig5}, t$_{\text{init}}$ = 2 and 9 $\mu$s produce different $\rho^{-}(t)$ (Fig.~\ref{fig4}), and thus fitting to $\exp(-\Gamma t)$ results in different values of $\Gamma$, neither of which are $\Gamma_1$. In practice, the bi-exponential decay of PL$_{\text{diff},T_1}$ may be nearly indistinguishable from a mono-exponential decay with $\Gamma \approx \Gamma_1 + A \Gamma_c$, and so we emphasize that NV$^-$ ionization in the dark requires attention. 

To fully mitigate charge ionization in the dark, in Fig.~\ref{fig5}(b) we normalize PL$_{\text{diff},T_1}$ by $\rho^{-}(t)$, which is measured separately in the same measurement sequence. We recover a $T_1$ decay free from the effects of ionization in the dark: the data and the fitted values of $\Gamma$ agree for the two initialization times. The same analysis and mitigation protocol hold for $T_2$ measurements as well. See SI Note 4 \cite{Note1} for a discussion of other cases and the corresponding mitigation protocols.

In conclusion, we show that the charge state properties of NV centers both under illumination and in the dark depend on the charge configuration of the local environment, and shallow NV centers can exhibit a significantly lower and less stable NV$^{-}$ population relative to bulk NVs. These observations have direct implications in measurement sensitivity and validity, which can be addressed with the various measurement protocols we present here. We also identify the origin of NV$^{-}$ ionization in the dark as tunneling to a single local electron trap; we achieve control and readout of the trap charge state and measure its optical ionization properties. Future experiments can use these control and readout capabilities to directly identify the trap's structure and characterize the NV-trap tunneling mechanism \cite{Chou2018}. For instance, one can measure NV-trap separation by measuring trap-state-dependent electric fields \cite{Dolde2014,Mittiga2018}, or one could measure the trap position in the bandgap by varying the optical excitation energy. On the other hand, the NV-trap tunneling mechanism could be utilized for quantitative and highly sensitive measurements of electrochemical potentials, or for the production of tunable local electric and magnetic fields for sensing applications, among other potential applications.

\begin{acknowledgments}
We thank Claire McLellan, Amila Ariyaratne, Nathalie de Leon, Norman Yao, Patrice Bertet, and Chris Van de Walle for helpful discussions. We acknowledge partial support by a PECASE award from the Air Force Office of Scientific Research and partial support from NSF CAREER Grant No. DMR-1352660. D.B. acknowledges funding from the Microscopy Society of America and the Barry Goldwater Foundation.
\end{acknowledgments}

\bibliography{Charge_state_paper.bib}

\clearpage
\onecolumngrid
\subsection*{\Large Supplementary Information}
\normalsize

\setcounter{equation}{0}
\setcounter{figure}{0}
\setcounter{table}{0}
\makeatletter
\renewcommand{\theequation}{S\arabic{equation}}
\renewcommand{\thefigure}{S\arabic{figure}}
\renewcommand{\thetable}{S\arabic{table}}

\section{Supplementary note 1: Sample preparation and details}

The diamond is prepared by growth of a 50-nm thick 99.99\% $^{12}$C isotopically purified film onto a commercial Element 6 electronic grade (100) diamond substrate. Before growth, the diamond is etched by ArCl$_2$ plasma (1 $\mu$m) to alleviate polishing damage and cleaned in boiling acid H$_2$NO$_3$:H$_2$SO$_4$ 2:3. NV centers are formed by $^{14}$N ion implantation at 4 keV and a 7$\degree$ tilt with a dosage of $5.2\times 10^{10}$ ions / cm$^2$. These implantation parameters yield an expected N depth of 7 nm as calculated by Stopping and Range of Ions in Matter (SRIM). The sample is then annealed in vacuum ($< 10^{-6}$ Torr) at 850$\degree$ C for 2.5 hours with a 40-minute temperature ramp. After the annealing step, the diamond is cleaned in HClO$_4$:H$_2$NO$_3$:H$_2$SO$_4$ 1:1:1 for 1 hour at 230-240 $\degree$C.

A waveguide patterned on the diamond is used to transmit microwaves to coherently drive transitions between the spin states of the NV$^-$ ground state triplet. Optical access in the confocal setup is through the 150-$\mu$m thick diamond plate. To enhance photon collection efficiency, tapered nanopillars with a diameter of 400-nm are patterned with e-beam lithography and etched in O$_2$ plasma to a height of 500 nm. No NV centers are observed between the 500-nm-tall etched pillars, indicating that the density of NVs from native N centers is negligible; as such all measured NVs are assumed to be implanted. Pillars with single NV centers are found with second-order correlation measurements and confirmed with charge state photon statistics (see SI Note 5). The saturation power of the NV fluorescence is measured to be $\sim$ 200-$\mu$W under CW 532-nm excitation. The NV depth is measured with proton nuclear magnetic resonance (see Fig.~\ref{N3depth}) and ranges between $\sim$ 3-17 nm.

Several rounds of surface cleaning are performed between the initial sample preparation and the measurements performed in this work. The measurements plotted in the main text are performed in the months after a standard surface preparation protocol consisting of an acid clean and oxygen anneal. The acid clean consists of 12 minutes of Nanostrip (a Piranha analog) at 80$^{\circ}$C and the oxygen anneal is performed at 400$^{\circ}$C for 4 hours. Nanostrip is chosen to preserve the metal waveguide.

\section{Supplementary note 2: Charge state readout}

\subsection*{2.1: Photon statistics model}

We perform charge state readout by collecting photons for $\sim$ 1 ms under $\sim$ 2-$\mu$W CW 594-nm excitation. As in Fig.~1 of the main text, we plot the probability to get n photons ${\rm P}(n)$ and then fit the statistics to the sum of two distributions, specifically

\begin{equation}
{\rm P}(n) = {\rm P}(n|{\rm NV}^-) {\rm P}({\rm NV}^-) + {\rm P}(n|{\rm NV}^0) {\rm P}({\rm NV}^0)
\label{pofn}
\end{equation}

where ${\rm P}({\rm NV}^-)$, ${\rm P}({\rm NV}^0)$ are the respective probabilities to be in NV$^-$, NV$^0$ immediately before readout, and ${\rm P}(n|{\rm NV}^-)$, ${\rm P}(n|{\rm NV}^0)$ are the respective probabilities to get $n$ photons given that the NV was in NV$^-$, NV$^0$ immediately before readout. Knowledge of the distributions ${\rm P}(n|{\rm NV}^-)$ and ${\rm P}(n|{\rm NV}^0)$ then allows to extract ${\rm P}({\rm NV}^-)$ and ${\rm P}({\rm NV}^0) = 1 - {\rm P}({\rm NV}^-)$.

If the NV charge state were stable during the entire readout period, then ${\rm P}(n|{\rm NV}^-) = {\rm PoissPDF}(\gamma_- t_R,n)$ and ${\rm P}(n|{\rm NV}^0) = {\rm PoissPDF}(\gamma_0 t_R,n)$, where $t_R$ is the total readout time, $\gamma_-$, $\gamma_0$ are the count rates from NV$^-$, NV$^0$, and PoissPDF$(\lambda,n)$ is the probability of an outcome $n$ for a Poisson random variable with mean value $\lambda$. 

However, the NV photoionization rate $g_{-0}$ and photorecombination rate $g_{0-}$ are non-negligible, and so to accurately calculate ${\rm P}(n|{\rm NV}^-)$ and ${\rm P}(n|{\rm NV}^0)$ one must account for the possibility of ionization and recombination during readout. For instance, if the NV starts in NV$^-$ and then ionizes to NV$^0$ halfway through the readout, then ${\rm P}(n) = {\rm PoissPDF}(\gamma_- t_R/2 + \gamma_0 t_R/2,n)$. The appropriate ${\rm P}(n|{\rm NV}^-)$ and ${\rm P}(n|{\rm NV}^0)$ are thus sums over an an infinite number of Poisson distributions, weighted by the probability for each ionization sequence to occur. This weighted infinite sum is calculated by Shields \textit{et al.} \cite{Shields2015} and Hacquebard \textit{et al.} \cite{Hacquebard2018} to arrive at

\begin{equation}
{\rm P}(n|{\rm NV}^-,{\rm odd}) = \int_0^{t_R}d\tau g_{-0} e^{(g_{0-}-g_{-0})\tau - g_{0-}t_R} \text{BesselI}(0,2\sqrt{g_{-0}g_{0-}\tau(t_R-\tau)}) {\rm PoissPDF}(\gamma_-\tau+\gamma_0(t_R-\tau),n)
\end{equation}
\begin{equation}
\begin{aligned}
{\rm P}(n|{\rm NV}^-,{\rm even}) =& \int_0^{t_R}d\tau \sqrt[]{\frac{g_{-0}g_{0-}\tau}{t_R-\tau}} e^{(g_{0-}-g_{-0})\tau - g_{0-}t_R} \text{BesselI}(1,2\sqrt{g_{-0}g_{0-}\tau(t_R-\tau)}) {\rm PoissPDF}(\gamma_-\tau+\gamma_0(t_R-\tau),n)\\
&+e^{-g_{-0}t_R}{\rm PoissPDF}(\gamma_- t_R,n)
\end{aligned}
\end{equation}
\begin{equation}
{\rm P}(n|{\rm NV}^-) = {\rm P}(n|{\rm NV}^-,{\rm odd}) + {\rm P}(n|{\rm NV}^-,{\rm even})
\end{equation}

where BesselI$(m,x)$ is a modified Bessel function of the first kind. To calculate ${\rm P}(n|{\rm NV}^0)$ one can simply exchange the subscripts $- \leftrightarrow 0$.

Under CW excitation, Eq.~(\ref{pofn}) can be rewritten with the ionization rate $g_{-0}$ and recombination rate $g_{0-}$ because ${\rm P}({\rm NV}^-) / {\rm P}({\rm NV}^0) = g_{0-} / g_{-0}$ in the steady state. So, in that case

\begin{equation}
{\rm P}(n) = {\rm P}(n|{\rm NV}^-) \left(\frac{1}{1+g_{-0}/g_{0-}}\right) + {\rm P}(n|{\rm NV}^0) \left(\frac{1}{1+g_{0-}/g_{-0}}\right)
\label{Cwfit}
\end{equation}

As in Refs \cite{Shields2015} and \cite{Hacquebard2018} we bin photons in $\sim$ ms windows and fit to the statistics using Eq.~(\ref{Cwfit}), thus extracting $\gamma_-, \gamma_0, g_{-0},$ and $g_{0-}$ as 4 free fit parameters. In practice, during the measurement we bin the photons in $100$-$\mu$s windows and then, in postprocessing, increase the bin size in multiples of $100$-$\mu$s. We find that this variable bin size is an effective way to verify the validity of the fit.

\subsection*{2.2: Measuring $\rho^{-}$}

In this work, however, we are typically interested in ${\rm P}({\rm NV}^-)$ immediately after green excitation or in the dark. In this case we treat ${\rm P}({\rm NV}^-)$ as a free fit parameter in Eq.~(\ref{pofn}), as in Figs.~1(c) and 1(d) of the main text, in addition to the 4 free fit parameters $\gamma_-, \gamma_0, g_{-0},$ and $g_{0-}$ (see above). However, this fit cannot always be reasonably carried out; for instance, in Fig.~1(f) of the main text we calculate $\rho^- \equiv {\rm P}({\rm NV}^-)$ for sets of only 1000 readouts, which is far insufficient to accumulate significant statistics as in Figs.~1(c) and 1(d). Accordingly, to calculate $\rho^-$ in Fig.~1(f), we fit the distributions ${\rm P}(n|{\rm NV}^-)$ and ${\rm P}(n|{\rm NV}^0)$ to the data in Fig.~1(d) and then explicitly calculate

\begin{align}
\langle n_- \rangle \equiv \sum_{n=0}^\infty n {\rm P}(n|{\rm NV}^-) && \langle n_0 \rangle \equiv \sum_{n=0}^\infty n {\rm P}(n|{\rm NV}^0)
\label{navg}
\end{align}

\begin{equation}
\langle n \rangle = \rho^- \langle n_- \rangle + \rho^0 \langle n_0 \rangle
\end{equation}

\begin{equation}
\rho^- = \frac{\langle n \rangle - \langle n_0 \rangle}{\langle n_- \rangle - \langle n_0 \rangle}
\label{rhocalc}
\end{equation}

\begin{figure}
\includegraphics[width=178mm]{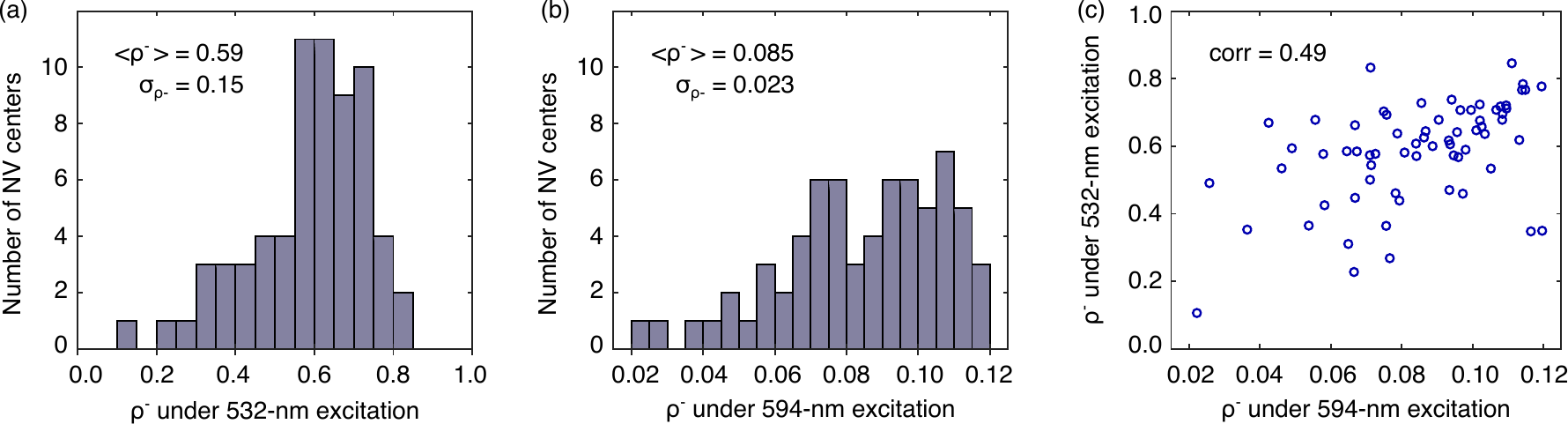}
\caption{Average NV$^-$ population $\rho^-$ under CW illumination for 67 individual centers. (a) $\rho^-$ under CW, 300-$\mu$W 532-nm excitation, measured with the sequence presented in Fig.~1 of the main text. (b) $\rho^-$ under CW, 2-$\mu$W 594-nm excitation, measured by fitting the photon statistics with the method explained in Eq.~(\ref{Cwfit}). (c) Observed correlation between data in (a) and (b), with a Pearson correlation coefficient of 0.49. For bulk NV centers, the typical reported value of $\rho^-$ is $\approx$ 0.75 under CW 532-nm excitation \cite{Aslam2013,Shields2015,Doi2016,Doi2014,Hopper2016,Chen2013} and $\approx$ 0.12 under CW 594-nm excitation \cite{Aslam2013,Shields2015,Hacquebard2018,Doi2016,Hopper2016}.}
\label{NVstat}
\end{figure}

where $\langle n \rangle$ is the measured mean number of photons per shot, $\rho^0 = 1 - \rho^-$, and the sums in Eq.~(\ref{navg}) are approximated by summing until $n = 10 \gamma_- t_R$ instead of $n = \infty$. We use this method of explicit calculation to calculate $\rho^-$ and $\rho^-(t)$ throughout the paper, except in Figs.~1(c) and 1(d) of the main text. We typically calibrate ${\rm P}(n|{\rm NV}^-)$ and ${\rm P}(n|{\rm NV}^0)$ with CW 594-nm excitation to reduce the number of free fit parameters, as explained in Eq.~(\ref{Cwfit}). However, we find that on some shallow NVs the photoionization rate $g_{-0}$ can increase after a high-power 532-nm pulse, and this can affect ${\rm P}(n|{\rm NV}^-)$ and ${\rm P}(n|{\rm NV}^0)$. We attribute this effect to creation of acceptors by the high-power pulse. Accordingly, we use a readout time $t_R < 1/g_{-0} \ll 1/g_{0-}$ so that the distributions and Eq.~(\ref{navg}) are only weakly dependent on $g_{-0}$ and $g_{0-}$.

\subsection{2.3: NV $\rho^-$ statistics}

In Fig.~\ref{NVstat} we plot the average NV$^-$ population $\rho^-$ with 532-nm and 594-nm illumination for 67 individual NV centers in the same sample. In Fig.~\ref{NVstat}(a) we measure $\rho^-$ under CW, 300-$\mu$W 532-nm excitation with the sequence presented in Fig.~1 of the main text. In Fig.~\ref{NVstat}(b) we measure $\rho^-$ under CW 2-$\mu$W 594-nm excitation by fitting the photon statistics with the method explained in Eq.~(\ref{Cwfit}). To measure $\rho^-$ in Fig.~\ref{NVstat}(a) we directly apply Eq.~(\ref{rhocalc}) with the fitted photon distributions ${\rm P}(n|{\rm NV}^-)$ and ${\rm P}(n|{\rm NV}^0)$ from Fig.~\ref{NVstat}(b). We warn that this method of calculation is not as quantitatively reliable as explicitly fitting the photon statistics, as performed in Figs.~1(c) and 1(d) of the main text. We also emphasize the observation that under both 532-nm and 594-nm excitation $\rho^-$ gradually decreases over time, and so the distributions in Fig.~\ref{NVstat} will gradually shift leftward. All 67 centers presented in Fig.~\ref{NVstat} were measured over a span of 11 days, several months after a surface cleaning consisting of a 10 minute Piranha ($\text{H}_2\text{SO}_4$:$\text{H}_2\text{O}_2$ 3:1) boil.

\section{Supplementary note 3: Evidence of a single electron trap}
\subsection{3.1: Non-mono-exponential behavior from multiple traps}

In this section we prove that the data presented in the main text for the NV$^{-}$ survival probability, P(NV$^{-}$ not decay) = $\rho^{-}(t) / \rho^{-}(0)$, can only be explained by a single dominant electron trap.

Consider electron tunneling from the NV at a constant rate $\Gamma_c$ to an empty single electron trap. In this case, P(NV$^{-}$ not decay $|$ trap empty) = $\exp\left(-\Gamma_c t\right)$. However, if the trap is full, then P(NV$^{-}$ not decay $|$ trap full) = 1. If P(trap empty) = $A$, then P(trap full) = $1 - A$ and 

\begin{equation}
\begin{split}
\text{$\rho^{-}(t) / \rho^{-}(0)$ = P(NV$^{-}$ not decay)} = (1 - A) + A e^{-\Gamma_c t}
\\
\rho^{-}(t) / \rho^{-}(0) = 1 - A \left(1-e^{-\Gamma_c t} \right)
\label{fitmodelS}
\end{split}
\end{equation}

Consider instead that there are N electron traps indexed by $i$, in which case our model predicts that P(not decay to $i^{\text{th}}$ trap) =  $1 - A_i \left(1-e^{-\Gamma_i t} \right)$ where $A_i$ is the probability for the $i^{\text{th}}$ trap to be empty and $\Gamma_i$ is the NV tunneling rate to the $i^{\text{th}}$ trap. In this case

\begin{equation}
\rho^{-}(t) / \rho^{-}(0) = \text{P(not decay) = $\Pi_i$P(not decay to $i^{\text{th}}$ trap)} = \Pi_i \left(1 - A_i \left(1-e^{-\Gamma_i t} \right)\right)
\label{Ntraps}
\end{equation}

This product necessarily produces non-mono-exponential decays unless $A_i = 1 ~\forall i$. In the main text we observe $\rho^{-}(t) / \rho^{-}(0)$ to fit well to a mono-exponential decay to a nonzero value $A$, as given by Eq.~(\ref{fitmodelS}), thereby suggesting a single dominant electron trap. 

However, non-mono-exponential decays can closely resemble mono-exponential decays. Nevertheless, with a non-mono-exponential decay the fitted $\Gamma_c$ in Eq.~(\ref{fitmodelS}) would necessarily increase as the fitted $A$ increases from 0 to 1, as we prove in the following sections. This signature of non-mono-exponential decays then serves as a strong metric for the presence of multiple dominant traps because we can vary $A$ experimentally by varying the initialization time and power (main text Fig.~4 and Fig.~\ref{GammaT}). We then show that our fitted $\Gamma_c$ does not increase as $A$ increases, thereby demonstrating that our NV is sensitive to only a single dominant trap.

\subsection*{3.2: $N = 2$ electron traps}

We first show that for $N = 2$ dominant electron traps the fitted $\Gamma_c$ in Eq.~(\ref{fitmodelS}) will increase as the fitted $A$ increases. In main text Fig.~4 we experimentally vary $A$ from $\sim$ 0.2 to 1. Accordingly, in the following mathematical analysis we consider the limiting cases of $A \ll 1$ and $A = 1$ to predict how $\Gamma_c$ would scale over the experimentally observed range of $A$. Here we define two dominant traps by $\Gamma_1 \approx \Gamma_2$ and $A_1 \approx A_2$.

Using Eq.~(\ref{Ntraps}) we see that for two traps we have

\begin{equation}
\begin{split}
\rho^{-}(t) / \rho^{-}(0) &= \left(1 - A_1 \left(1-e^{-\Gamma_1 t}\right)\right)*\left(1 - A_2 \left(1-e^{-\Gamma_2 t}\right)\right)
\\
&= (1-A_1)(1-A_2) + A_1(1-A_2)e^{-\Gamma_1 t} + A_2(1-A_1)e^{-\Gamma_2 t} + A_1 A_2 e^{-\left(\Gamma_1 + \Gamma_2\right)t}
\end{split}
\end{equation}

In the case where this expression resembles Eq.~(\ref{fitmodelS}) with $A \ll 1$, then $A_1, A_2 \ll 1$. We can then approximate to 

\begin{equation}
\begin{split}
\rho^{-}(t) / \rho^{-}(0) &\approx \left(1 - A_1 - A_2\right) + \left(A_1e^{-\Gamma_1 t} + A_2e^{-\Gamma_2 t}\right)
\\
&\approx \left(1 - \left(A_1 + A_2\right)\right) + \left(A_1 + A_2\right) \exp\left(-\left(\frac{A_1 \Gamma_1 + A_2 \Gamma_2}{A_1 + A_2}\right)t\right)
\end{split}
\end{equation}

So with $A \ll 1$, we recover the form of the mono-exponential decay in Eq.~(\ref{fitmodelS}) with $A = A_1 + A_2$ and $\Gamma_c = \left(A_1 \Gamma_1 + A_2 \Gamma_2\right) / \left(A_1 + A_2\right)$.

In contrast, when $A = 1$ then $A_1 = A_2 = 1$ (under our definition that dominant traps have $A_1 \approx A_2$). In this case we again recover a mono-exponential decay

\begin{equation}
\rho^{-}(t) / \rho^{-}(0) = e^{-\left(\Gamma_1 + \Gamma_2\right) t}
\end{equation}

This resembles Eq.~(\ref{fitmodelS}) with $A = 1$ and $\Gamma_c = \Gamma_1 + \Gamma_2$. So we see that in the case of two dominant traps, the apparent $\Gamma_c$ increases as $A$ increases from 0 to 1. Specifically, since $\Gamma_1 \approx \Gamma_2$ and $A_1 \approx A_2$, $\Gamma_c$ approximately doubles from $\left(\Gamma_1 + \Gamma_2\right)/2$ to $\Gamma_1 + \Gamma_2$. Note that if $A_1 \sim A_2$ but $\Gamma_1$ is not approximately $\Gamma_2$, then the data would not fit well to a mono-exponential decay unless $\Gamma_1 \gg \Gamma_2$ (or vice versa) which is the case of only one dominant trap.

The same analysis holds for all cases of $N \sim 1$ traps; for instance, for N = 3 traps the apparent $\Gamma_c$ triples as $A$ increases from 0 to 1. In the next section we analyze the case of $N \gg 1$ traps.

\subsection*{3.3: $N \gg 1$ electron traps}

Here we show that for $N \gg 1$ dominant electron traps the apparent $\Gamma_c$ in Eq.~(\ref{fitmodelS}) will increase with $A$. For simplicity we assume $A_i = a$ and $\Gamma_i = \gamma$ for all the dominant traps. This turns Eq.~(\ref{Ntraps}) to

\begin{equation}
\rho^{-}(t) / \rho^{-}(0) = \left(1 - a \left(1 -  e^{-\gamma t} \right)\right)^N
\end{equation}

If $A \ll 1$ then $a \ll 1$. In this case we Taylor expand to find

\begin{equation}
\rho^{-}(t) / \rho^{-}(0) \approx 1 - N a \left(1 -  e^{-\gamma t}\right)
\end{equation}

So with $A \ll 1$ we recover the form of the mono-exponential decay in Eq.~(\ref{fitmodelS}) with $A = Na$ and $\Gamma_c = \gamma$. With $A = 1$ we again want $\rho^{-}(t) / \rho^{-}(0)$ to resemble $e^{-\Gamma_c t}$
\begin{equation}
\begin{split}
\rho^{-}(t) / \rho^{-}(0) = \left(1 - a \left(1 -  e^{-\gamma t} \right)\right)^N = e^{-\Gamma_c t}
\\
1 - a \left(1 -  e^{-\gamma t} \right) = e^{-\Gamma_c t / N}
\end{split}
\end{equation}

But $\Gamma_c t \sim 1$ for the $\tau$ points measured, so $\Gamma_c t / N \ll 1$ and $\gamma t \ll 1$. So we Taylor expand and find

\begin{equation}
\begin{split}
1 - a \gamma t = 1 - \Gamma_c t / N
\\
\Gamma_c = N a \gamma
\end{split}
\end{equation}

With $Na \gg 1$ because $\gamma t \ll 1$ and $\Gamma_c t \sim 1$ for the $\tau$ points measured. 

So, we find that as $A$ increases from 0 to 1, for N traps with $A_i = a$ and $\Gamma_i = \gamma$, the apparent $\Gamma_c$ increases from $\gamma$ to $Na\gamma ~\text{with}~ Na \gg 1$. If $a = 1$, then we recover the result of the previous section that the apparent $\Gamma_c$ increases from $\gamma$ to $N \gamma$.

\subsection*{3.4: $\Gamma_c$ vs. $A$ data as evidence of a single dominant trap}

\begin{figure}
\includegraphics[width=175mm]{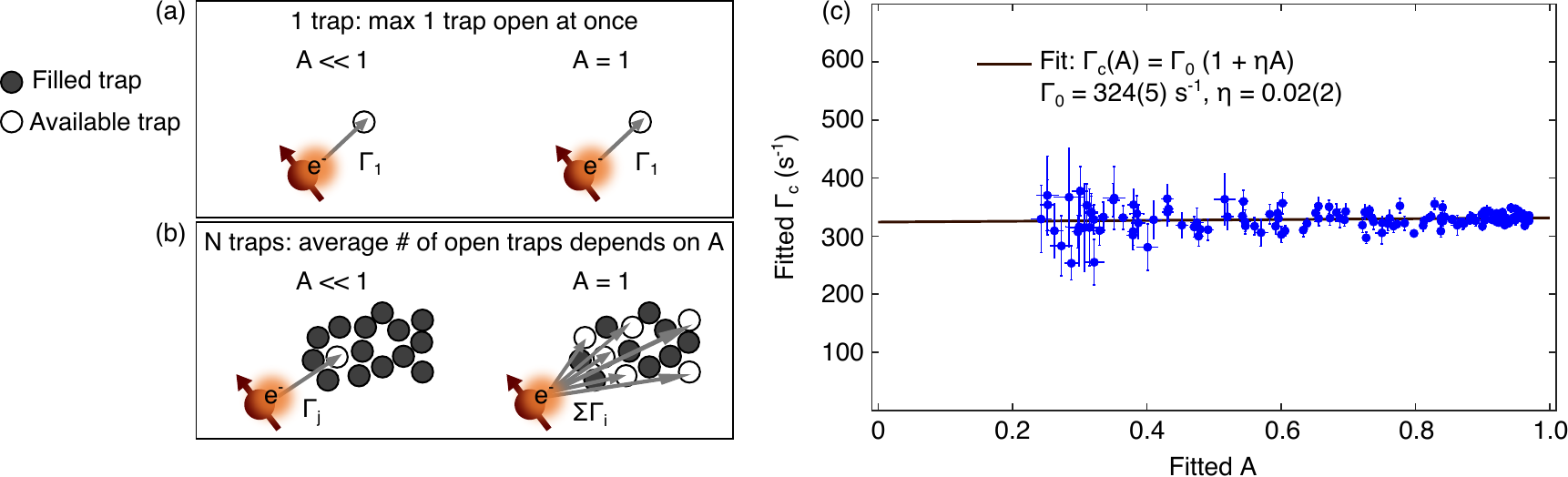}
\caption{The number of dominant electron traps affects the effective NV$^-$ decay rate $\Gamma_c$ versus the probability to decay $A$. (a) Schematic of NV tunneling to a single trap. $\rho^{-}(t) / \rho^{-}(0)$ will decay at rate $\Gamma_c = \Gamma_1$ to a finite value $A$ [Eq.~(\ref{fitmodelS})] regardless of whether $A \ll 1$ or $A = 1$. (b) Schematic of NV tunneling to multiple traps. When $A \ll 1$ there will statistically only be one trap open at a time and $\Gamma_c \approx \Gamma_j$ where the j$^{th}$ trap is empty. However with $A = 1$, there will statistically be multiple traps open at a time and $\Gamma_c \approx \Sigma_i \Gamma_i$ where $i$ indexes over the open traps. (c) Fitted $\Gamma_c$ versus fitted $A$ from the 135 data points in Fig.~4 of the main text, where $A$ is varied experimentally by varying initialization time and power. We interpolate $\Gamma_c(A) = \Gamma_0 \left(1 + \eta A\right)$ and fit $\eta = 0.02(2)$. If there is more than one dominant trap then $\eta \gtrsim 1$, and so this NV is only sensitive to a single dominant electron trap. Since $A$ varies with initialization time and power, this plot equivalently shows $\Gamma_c$ does not depend on initialization time or power.}
\label{GammaVsA}
\end{figure}

In the previous section we show that the distinguishing signature of a single dominant trap is a mono-exponential decay of $\rho^{-}(t) / \rho^{-}(0)$ to a nonzero value $A$. However, distinguishing between mono-exponential and non-mono-exponential decays is challenging, and so we also show in the previous sections that a practical metric of a single dominant electron trap is a fitted $\Gamma_c$ which does not increase with $A$. For $N>1$ dominant traps, $\Gamma_c$ increases the least for $N = 2$, approximately doubling as $A$ increases from 0 to 1.

In Fig.~\ref{GammaVsA} we plot the fitted $\Gamma_c$ vs the fitted $A$ from the data in Fig.~4 of the main text, where we vary $A$ experimentally by varying initialization time and power. We interpolate $\Gamma_c(A) = \Gamma_0 \left(1 + \eta A\right)$ and fit $\eta = 0.02(2)$. If there is more than one dominant trap then $\eta \gtrsim 1$, and so this NV is only sensitive to a single dominant electron trap.

\section{Supplementary note 4: Spin measurements in the presence of charge conversion in the dark}
\subsection{4.1: Spin evolution in the presence of charge conversion}
In this section we show that charge conversion in the dark can compromise the validity of NV spin measurement signals such as $T_1$ and $T_2$ decay measurements. In general, the spin populations are functions of time due to spin processes such as $T_1$ relaxation. However, in the presence of charge conversion in the dark, the NV$^-$ and NV$^0$ populations also become functions of time, which introduces additional time-dependent PL signatures. These detrimental effects are mitigated by the measurement protocols presented here. 

Mathematically, a photoluminescence (PL) readout yields

\begin{flalign}
\text{PL}(t) = \text{PL}^-_0 \rho^-_0(t) + \text{PL}^-_{+1} \rho^-_{+1}(t) + \text{PL}^-_{-1} \rho^-_{-1}(t) + \text{PL}^0 \rho^0(t)
\label{PLsignal}
\end{flalign}

where $\rho^-_0, \rho^-_{+1}, \rho^-_{-1}$ are the respective populations of the $m_s = 0,+1,-1$ spin states of NV$^-$, $\rho^0$ is the NV$^0$ population, and $\text{PL}^-_0, \text{PL}^-_{+1},\text{PL}^-_{-1},\text{PL}^0$ describe the PL from the various states. We note that $\text{PL}^-_{+1} = \text{PL}^-_{-1}$.

We now consider the effect of charge conversion on a $T_1$ measurement signal as an example. In a $T_1$ measurement we initialize optically into $m_s = 0$ and then let the spin relax in the dark. As derived in the supplements of \cite{Myers2017a} and \cite{Ariyaratne2018a} in the absence of charge conversion, the spin populations evolve in time with

\begin{align}
\text{No charge conversion in the dark:} \begin{cases}
\rho^-_0(t) = \left(1/3 + 2/3~ \exp\left(-\Gamma_1 t\right)\right) \rho^-(0)
\\
\rho^-_{\pm1}(t) = \left(1/3 - 1/3 ~\exp\left(-\Gamma_1 t\right)\right) \rho^-(0)
\end{cases}
\end{align}

where $\rho^-(0)$ is the NV$^-$ population and here the populations are defined such that $\rho^-_0(t) + \rho^-_{-1}(t) + \rho^-_{+1}(t) = \rho^-(0)$, and $\Gamma_1 \equiv 1/T_1$. In practice, the optical spin polarization into $m_s = 0$ is non-unity, as described in \cite{Myers2017a} and \cite{Ariyaratne2018a}, but for brevity we assume perfect initial spin polarization and note that imperfect polarization merely reduces the apparent spin-state contrast for a differential measurement (described later). In the case of charge conversion in the dark, these populations then evolve in time with

\begin{align}
\text{Spin-independent ionization in the dark:} \begin{cases}
\rho^-_0(t) = \left(1/3 + 2/3~ \exp\left(-\Gamma_1 t\right)\right) \rho^-(t)
\\
\rho^-_{\pm1}(t) = \left(1/3 - 1/3 ~\exp\left(-\Gamma_1 t\right)\right) \rho^-(t)
\label{PopIon}
\end{cases}
\end{align}

\begin{align}
\text{Spin-independent recombination in the dark:} \begin{cases}
\rho^-_0(t) = \left(1/3 + 2/3 ~\exp\left(-\Gamma_1 t\right)\right) \rho^-(0) + 1/3 ~\left(\rho^-(t) - \rho^-(0)\right)
\\
\rho^-_{\pm1}(t) = \left(1/3 - 1/3~ \exp\left(-\Gamma_1 t\right)\right) \rho^-(0) + 1/3 ~\left(\rho^-(t) - \rho^-(0)\right)
\label{PopRecomb}
\end{cases}
\end{align}

In both cases $\rho^-_0(t) + \rho^-_{-1}(t) + \rho^-_{+1}(t) = \rho^-(t)$. However, in the case of spin-independent ionization the populations all gain a multiplicative time dependence, whereas in the case of spin-independent recombination the populations all gain an additive time dependence. The presence of additional time dependences motivates the use of common-mode rejection techniques, such as the differential measurement described below.

\subsection{4.2: Differential measurement and other normalization protocols}
A differential measurement \cite{Myers2017a} will alleviate the additive effects of recombination in the dark, but will not remove the multiplicative effects of ionization in the dark, which must be normalized with additional methods. Here we discuss the various possible cases and derive the corresponding mitigation protocols.

For a differential measurement, we perform readout at the end of the measurement sequence to measure PL as in Eq.~(\ref{PLsignal}). We then repeat the experiment but immediately before readout we apply a resonant microwave $\pi$ pulse to swap the $m_s=0$ and $m_s=1$ spin states to measure PL$_\text{swap}$ [see Fig.~\ref{OptT1}(a)]. This then yields the two signals

\begin{flalign}
\text{PL}(t) = \text{PL}^-_0 \rho^-_0(t) + \text{PL}^-_{+1} \rho^-_{+1}(t) + \text{PL}^-_{-1} \rho^-_{-1}(t) + \text{PL}^0 \rho^0(t)
\\
\text{PL$_\text{swap}$}(t) = \text{PL}^-_0 \rho^-_{+1}(t) + \text{PL}^-_{+1} \rho^-_{0}(t) + \text{PL}^-_{-1} \rho^-_{-1}(t) + \text{PL}^0 \rho^0(t)
\label{PLswap}
\end{flalign}

We then subtract the two signals to arrive at the differential measurement PL$_{\text{diff}}$

\begin{flalign}
\text{PL$_{\text{diff}}$}(t) = \text{PL}(t) - \text{PL$_\text{swap}$}(t) =  \left(\text{PL}^{-}_0 - \text{PL}^{-}_{+1}\right)\left(\rho^{-}_0(t) - \rho^{-}_{+1}(t)\right)
\label{diffsignal}
\end{flalign}

In the case of a $T_1$ measurement, we can use the population evolution in time expressed in Eq.~(\ref{PopIon}) and Eq.~(\ref{PopRecomb}) to arrive at 

\begin{flalign}
\text{Spin-independent ionization in the dark: PL$_{\text{diff},T_1}$}(t) = C \exp{\left(-\Gamma_1 t \right)} \left(\rho^{-}(t) / \rho^{-}(0)\right)
\label{DiffIon}
\end{flalign}

\begin{flalign}
\text{Spin-independent recombination in the dark:  PL$_{\text{diff},T_1}$}(t) = C \exp{\left(-\Gamma_1 t \right)}
\label{DiffRecomb}
\end{flalign}

where $C \equiv \left(\text{PL}^{-}_0 - \text{PL}^{-}_{+1}\right) \rho^-(0)$. In the case of spin-independent recombination in the dark, the populations gain the same additive time dependence, which is removed by the differential measurement. However, to recover a proper $T_1$ measurement from the multiplicative effects of ionization in the dark, one must separately measure and divide out the time dependence $\left(\rho^{-}(t) / \rho^{-}(0)\right)$, as demonstrated in Fig.~5 of the main text.

If both ionization and recombination are present, then the above protocols will not mitigate the effects of charge conversion. We discuss this case here and present the appropriate mitigation protocol. This possibility may be particularly relevant for measurements on NV ensembles, where both ionization and recombination in the dark have been observed \cite{Dhomkar2018a,Giri2018,Choi2017}. In the most general case, the spin-independent ionization will modulate the initially polarized populations by a multiplicative factor $\xi(t)$ and the spin-independent recombination will modulate the populations by an additive factor $\delta(t)$/3. In this general case,

\begin{flalign}
\rho^-_0(t) &= \rho'^{-}_0(t)\xi(t) + \delta(t)/3 \\
\rho^-_{+1}(t) &= \rho'^{-}_{+1}(t)\xi(t) + \delta(t)/3
\label{GeneralPop}
\end{flalign}

where we define $\rho'^{-}_0(t)$ and $\rho'^{-}_{+1}(t)$ as the spin populations in the absence of charge conversion. In the absence of charge conversion $\rho'^-_0(t) + \rho'^-_{-1}(t) + \rho'^-_{+1}(t) = \rho^-(0)$, and here $\rho^-_0(t) + \rho^-_{-1}(t) + \rho^-_{+1}(t) = \rho^-(0) \xi(t) + \delta(t)$. For this general case of both ionization and recombination, the differential signal becomes

\begin{flalign}
\text{PL$_{\text{diff}}$}(t) &= \left(\text{PL}^{-}_0 - \text{PL}^{-}_{+1}\right)\left(\rho'^{-}_0(t) - \rho'^{-}_{+1}(t)\right) \xi(t)
\\
\text{PL$_{\text{diff}}$}(t) &= \text{PL$'_{\text{diff}}$}(t) \xi(t)
\end{flalign}

where $\text{PL$'_{\text{diff}}$}(t)$ is the differential PL signal in the case that there is no charge conversion in the dark. This poses a challenge for $T_1$ measurements as one would have to determine $\xi(t)$. However, for measurements much shorter than $T_1$, $\exp(-\Gamma_1 t) \rightarrow 1$ in Eq.~(\ref{DiffIon}), which would now give $\text{PL$_{\text{diff,$T_1$}}$}(t) = C \xi(t)$. Thus, for measurements much shorter than $T_1$ one can normalize the measured $\text{PL$_{\text{diff}}$}(t)$ by $\text{PL$_{\text{diff,$T_1$}}$}(t)$. The differential measurement will remove the additive $\delta(t)$ and dividing by $\text{PL$_{\text{diff,$T_1$}}$}(t)$ will remove the multiplicative $\xi(t)$.

\subsection{4.3: Conclusion and discussion of charge conversion effects on spin measurements}
In conclusion of the previous sections, we have shown that the validity of spin measurements may be compromised by charge conversion in the dark, and we have presented measurement protocols to properly mitigate charge conversion in the various possible cases. To summarize the various cases:

\begin{description}
  \item[$\bullet$ Recombination in the dark] Perform differential measurement $\text{PL}_\text{diff}(t)$
  \item[$\bullet$ Ionization in the dark] Perform differential measurement $\text{PL}_\text{diff}(t)$ and then divide by $\rho^-(t)/\rho^-(0)$
  \item[$\bullet$ General case (ionization and/or recombination, or neither)] For a measurement much shorter than $T_1$, perform differential measurement $\text{PL}_\text{diff}(t)$ and then divide by $\text{PL}_\text{diff,$T_1$}(t)$
\end{description}

\begin{figure}
\includegraphics[width=178mm]{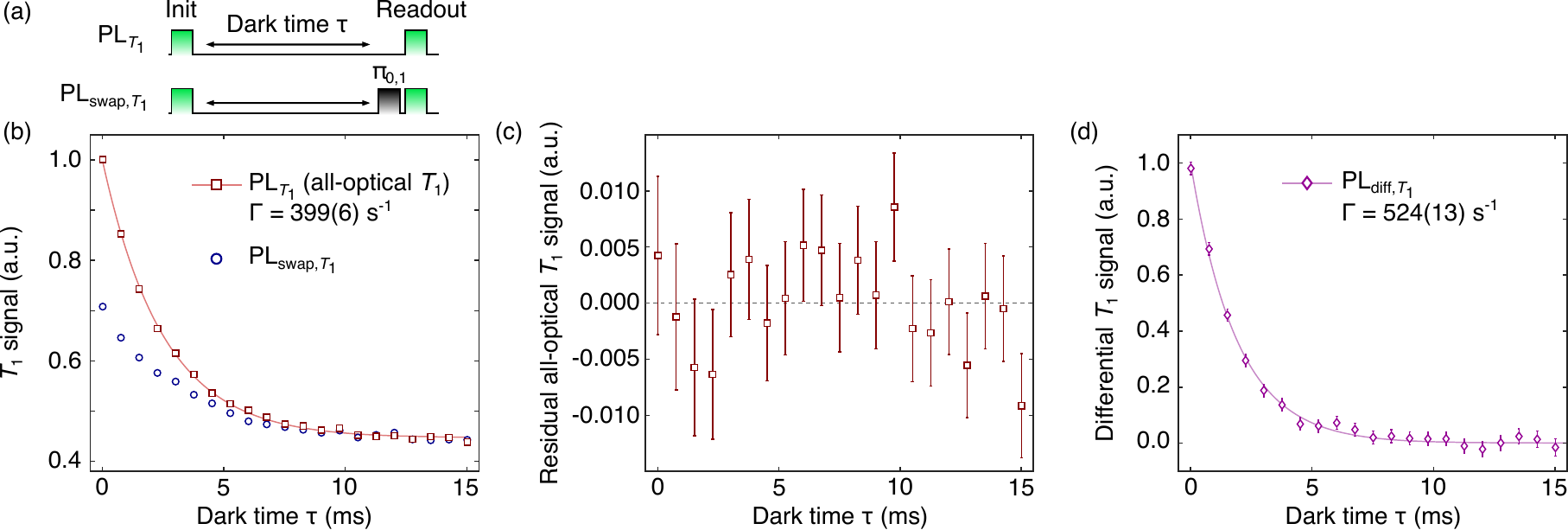}
\caption{Example $T_1$ signals in the presence of NV$^-$ ionization in the dark. (a) Pulse sequence to measure $\text{PL}_{T_1}$ and $\text{PL}_{\text{swap},T_1}$. (b) $\text{PL}_{T_1}$ and $\text{PL}_{\text{swap},T_1}$ signals measured on NV5 (same NV as Figs.~4-5 of the main text). Solid red curve is an exponential fit to the all-optical $T_1$ signal $\text{PL}_{T_1}$; due to the confounding effects of NV$^-$ ionization in the dark, the fit extracts an erroneous rate $\Gamma = 399(6) ~\text{s}^{-1} \neq \Gamma_1 = 182(5)~ \text{s}^{-1}$. Similarly, the $\text{PL}_{\text{swap},T_1}$ signal is predicted to increase with $\tau$, but instead is seen to decrease because of charge decay. (c) Residual of all-optical $T_1$ signal $\text{PL}_{T_1}$ from the exponential fit. Although the all-optical $T_1$ signal is not mono-exponential due to charge decay, the root-mean-square-residual is 0.004 and no clear deviation from the mono-exponential fit is observed. (d) Differential $T_1$ measurement $\text{PL}_{\text{diff,}T_1} = \text{PL}_{T_1} - \text{PL}_{\text{swap},T_1}$. Solid purple curve is a mono-exponential fit; due to the confounding effects of NV$^-$ ionization in the dark, the fit extracts an erroneous rate $\Gamma = 524(13) ~\text{s}^{-1} \neq \Gamma_1 = 182(5)~ \text{s}^{-1}$. The initialization power is $225 ~\mu$W and the initialization duration is $t_\text{init} = 200~ \mu$s.}
\label{OptT1}
\end{figure}

We now turn to a discussion of various signatures that could indicate the presence of charge conversion effects in spin measurements. In Fig.~\ref{OptT1}(b), we plot $\text{PL}_{T_1}$ and $\text{PL}_{\text{swap},T_1}$. In the absence of charge conversion, $\text{PL}_{\text{swap},T_1}$ is predicted to increase with $t$ (see \cite{Myers2017a}), but is clearly seen to decrease in Fig.~\ref{OptT1}(b), indicating the presence of confounding effects that obfuscate the real $T_1$ signal. In general, charge conversion effects may be seen in signals which decrease although they are predicted to increase (or vice versa), or signals which are non-monotonic although they are predicted to be monotonic. However, the all-optical $\text{PL}_{T_1}$ signal decays as expected and is well fit to a mono-exponential decay, not suggesting any issue with the measurement. We thus emphasize that these all-optical measurements are highly susceptible to undetected effects of charge conversion.

Nevertheless, the higher-than-expected signal contrast in $\text{PL}_{T_1}$ is an indication of charge conversion. The observed spin-state contrast (including imperfect spin and charge initialization) is 0.29, and without charge conversion $\text{PL}_{T_1}$ should decay to $1 - 2/3 * 0.29 = 0.81$ (see \cite{Myers2017a}), but instead is seen to decay to 0.45 because of charge conversion effects. We note, however, that these values depend on the relative values of $\text{PL}^-_0,\text{PL}^-_{\pm1}$, and $\text{PL}^0$, which depend on various parameters such as readout power and duration, excitation wavelength, and the spectral filters used in collection.

Initialization-dependence of a measurement signal is another possible indication of charge conversion effects. As shown in Fig.~4 of the main text, the charge conversion effect is highly initialization-dependent, with $A$ from Eq.~(\ref{fitmodelS}) increasing with initialization time and power. Accordingly, in Fig.~5 of the main text we observe that $\Gamma$ increases with initialization time, but that this effect disappears upon normalizing the signals by $\rho^-(t)/\rho^-(0)$. Similarly, for the all-optical $\text{PL}_{T_1}$ signal we observe that $\Gamma$ increases with initialization time (not plotted here).

\section{Supplementary note 5: Distinguishing single NV centers in nanostructures}

In order to distinguish a single photon emitter inside of a diffraction-limited spot, a second-order correlation measurement $g^{(2)}(\tau)$ is often employed, where $g^{(2)}(0) < 0.5$ is considered evidence of a single photon emitter. In this section we show that this is only true in the case where all possible emitters are equally bright. As such, we find $g^{(2)}(0) < 0.5$ is an insufficient criterion for distinguishing single NV centers in pillars, where the NV coupling to the waveguide modes is highly dependent on NV position. We confirm this effect by observing $g^{(2)}(0) < 0.5$ on an NV pillar that contains multiple NV orientations. We then show that the charge state photon statistics (as plotted in Fig.~1 of the main text) are highly sensitive to the presence of multiple NV centers, which we in turn use as a stronger criterion of distinguishing single NV centers.

We now derive the expected signal from a second-order correlation measurement $g^{(2)}(\tau)$ on two emitters with different brightnesses. To perform such a measurement, one places a beam splitter in the collection path to divert 50\% of the emission to detector 1 and 50\% to detector 2. With a coincidence counter (we use PicoHarp 300) one then measures

\begin{equation}
g^{(2)}(\tau) = \frac{\langle I_1(t) I_2 (t+\tau)\rangle}{\langle I_1(t) \rangle \langle I_2(t) \rangle}
\end{equation}

where $I_1(t)$ and $I_2(t)$ are the optical intensities at detectors 1 and 2 at time $t$, and $\langle \ldots \rangle$ denotes a time average. Suppose that we have two uncorrelated emitters that emit with intensities $\alpha(t) = \alpha_1(t) + \alpha_2(t)$ and $\beta(t) = \beta_1(t) + \beta_2(t)$ such that $I_1(t) = \alpha_1(t) + \beta_1(t)$ and $I_2(t) = \alpha_2(t) + \beta_2(t)$. Further, with the 50/50 beam splitter $\alpha \equiv \langle \alpha(t) \rangle = 2 \langle \alpha_1(t) \rangle = 2 \langle \alpha_2(t) \rangle$ and $\beta \equiv \langle \beta(t) \rangle = 2 \langle \beta_1(t) \rangle = 2 \langle \beta_2(t) \rangle$. So, we have then

\begin{equation}
\begin{split}
\langle I_1(t) \rangle \langle I_2(t) \rangle &= \langle \alpha_1(t) + \beta_1(t) \rangle \langle \alpha_2(t) + \beta_2(t) \rangle
\\
\langle I_1(t) \rangle \langle I_2(t) \rangle &= \left(\frac{\alpha + \beta}{2}\right)^2
\end{split}
\end{equation}

\begin{equation}
\begin{split}
\langle I_1(t) I_2(t+\tau) \rangle &= \langle \left(\alpha_1(t) + \beta_1(t)\right) \left(\alpha_2(t+\tau) + \beta_2(t+\tau)\right)\rangle
\\
&= \langle \alpha_1(t) \beta_2(t+\tau) \rangle + \langle \beta_1(t) \alpha_2(t+\tau) \rangle + \langle \alpha_1(t) \alpha_2(t+\tau) \rangle + \langle \beta_1(t) \beta_2(t+\tau) \rangle
\end{split}
\end{equation}

$\alpha_1(t)$ and $\beta_2(t)$ are uncorrelated so $\langle \alpha_1(t) \beta_2(t+\tau)\rangle = \langle \alpha_1(t) \rangle \langle \beta_2(t+\tau) \rangle = \alpha \beta / 4$. However, $\alpha_1(t)$ and $\alpha_2(t)$ are correlated because the emitters have a finite excited state lifetime $\tau_0$ which is the characteristic time required to emit another photon. Specifically, for an NV which emitted a photon at time $t$ (and so went to the ground state), the probability to emit another photon a time $\tau$ after $t$ is proportional to $\left(1 - \exp(-\tau / \tau_0)\right)$. So $\langle \alpha_1(t) \alpha_2(t+\tau) \rangle = (\alpha^2/4) * \left(1 - \exp(-\lvert\tau / \tau_0\rvert)\right)$. Finally,

\begin{equation}
\langle I_1(t) I_2(t+\tau) \rangle = \frac{\alpha \beta}{2} + \frac{\alpha^2 + \beta^2}{4} \left(1 - e^{-\lvert\tau/\tau_0\rvert}\right)
\end{equation}

\begin{gather}
g^{(2)}(\tau) = g^{(2)}(0) + \left(1 - g^{(2)}(0)\right)\left(1 - e^{-\lvert\tau/\tau_0\rvert}\right)
\label{g2fit}
\\
g^{(2)}(0) = \frac{2 \alpha / \beta}{\left(1 + \alpha / \beta\right)^2}
\label{g20}
\end{gather}

\begin{figure}
\includegraphics[width=162mm]{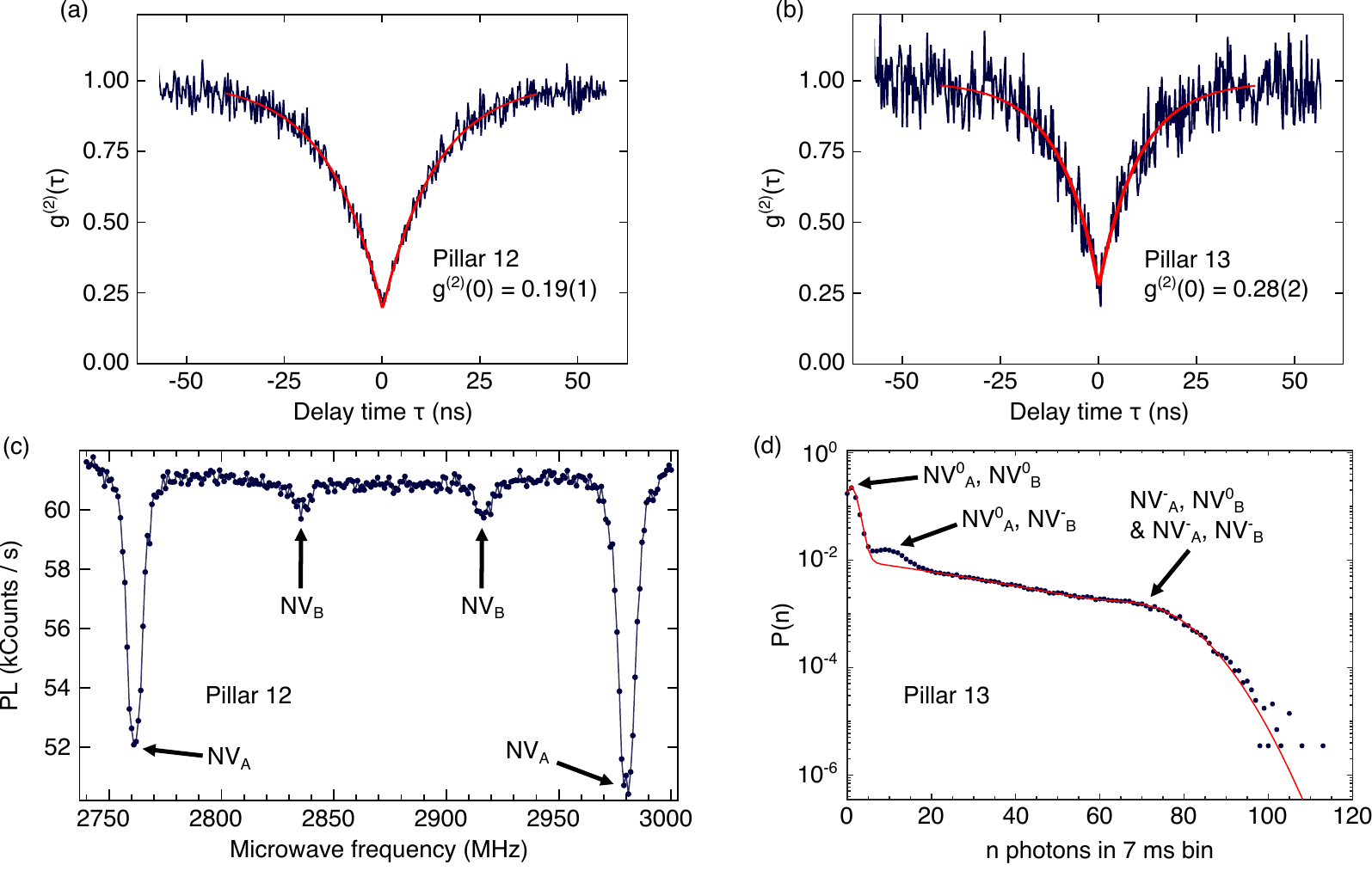}
\caption{Second-order correlation measurement $g^{(2)}(\tau)$ is insufficient to identify single NVs in nanostructures. (a),(b) $g^{(2)}(\tau)$ measurement on Pillar 12 and Pillar 13. Solid red curves are fits to Eq.~(\ref{g2fit}) and yield $g^{(2)}(0) = 0.19(1) < 0.5$ on Pillar 12 and $g^{(2)}(0) = 0.28(2) < 0.5$ on Pillar 13, suggesting that each pillar host only a single NV center. Laser power is chosen sufficiently low such that bunching shoulders from tangible shelving in the singlet state are not observed. (c) ODMR spectrum on Pillar 12 under continuous microwave and optical illumination, with an applied external magnetic field of $\sim$ 40-50 Gauss. Two sets of NV resonances are clearly observed, corresponding to the presence of two distinct NV orientations and accordingly at least two NV centers in Pillar 12. (d) Probability to measure $n$ photons $P(n)$ on Pillar 13 under $\sim 2$-$\mu$W CW 594-nm excitation, with photons binned in 7 ms windows. The data exhibit 3 peaks, which cannot be explained by the model (solid red curve) of a single NV hopping between NV$^-$ and NV$^0$ as presented in SI Note 2. However, the 3 peaks are well explained by the presence of two NV centers in Pillar 13, where NV$_\text{A}$ is approximately 7x brighter than NV$_\text{B}$.}
\label{M2andL7}
\end{figure}

In the case where the emitters are equally bright and $\alpha/\beta = 1$ we recover the result that $g^{(2)}(0) = 0.5$. However, if $\alpha / \beta \neq 1$ then $g^{(2)}(0) < 0.5$ may still be observed experimentally for sufficiently low dark counts and background. Note that the above derivation assumes 0 background signal from non-NV sources, which can be incorporated by considering a third source with Poissonian or Super-Poissonian characteristics.

In Fig.~\ref{M2andL7}(a) we plot the second-order correlation measurement for Pillar 12, and using Eq.~(\ref{g2fit}) we fit $g^{(2)}(0) = 0.19(1)$, suggesting that this pillar hosts only one NV center by the $g^{(2)}(0) < 0.5$ criterion. However, in Fig.~\ref{M2andL7}(c) we measure an optically detected magnetic resonance (ODMR) spectrum on the pillar and observe two sets of NV resonances, corresponding to two distinct NV orientations and at least two NV centers. The relative PL contrast of the two sets of resonances depends on a variety of parameters, such as the microwave and optical power. Approximately, the two resonances can be explained by two NV centers where NV$_\text{A}$ is 10x brighter than NV$_\text{B}$. Eq.~(\ref{g20}) gives $g^{(2)}(0) = 0.165$ for two NV centers where NV$_\text{A}$ is 10x brighter than NV$_\text{B}$, and accounting for the typical background signal observed in other NV pillars gives $g^{(2)}(0) \approx 0.20$, which agrees well with the fitted $g^{(2)}(0)$ in Fig.~\ref{M2andL7}(a).

Similarly, in Fig.~\ref{M2andL7}(b) we plot g$^{(2)}(\tau)$ on Pillar 13 and fit $g^{(2)}(0) = 0.28(2) < 0.5$. On this pillar we only observe one NV orientation in the ODMR spectrum (not plotted here), which is still consistent with a single center. However, the observed photon statistics under CW 594-nm excitation on Pillar 13 [Fig.~\ref{M2andL7}(d)] are inconsistent with the predicted photon statistics from a single NV center hopping between the bright NV$^-$ state and the dark NV$^0$ state. Instead, the statistics in Fig.~\ref{M2andL7}(d) suggest two NV centers where NV$_\text{A}$ is 7x brighter than NV$_\text{B}$. Eq.~(\ref{g20}) gives $g^{(2)}(0) = 0.22$ for two NV centers where NV$_\text{A}$ is 7x brighter than NV$_\text{B}$, and accounting for the typical background signal observed in other NV pillars gives $g^{(2)}(0) \approx 0.26$, which agrees well with the fitted $g^{(2)}(0)$ in Fig.~\ref{M2andL7}(b). These results demonstrate that the charge state photon statistics can be a valuable tool for verifying the presence of a single NV center. Although here we plot the photon statistics under CW 594-nm illumination, we find the 594-nm readout statistics after a 532-nm initialization pulse (as in Figs.~1(c) and 1(d) of the main text) can sometimes be more effective for discerning if there is more than one NV center.

To avoid artifacts from pillars with multiple NV centers, we use multiple criteria to find pillars with single centers: we measure $g^{(2)}(0)$, check for multiple NV orientations in the ODMR spectrum, and measure the charge state photon statistics. The NV centers used in this work pass all 3 tests.

\section{Supplementary Note 6: NV-based Nuclear magnetic resonance}

\begin{figure}
\includegraphics[width=91mm]{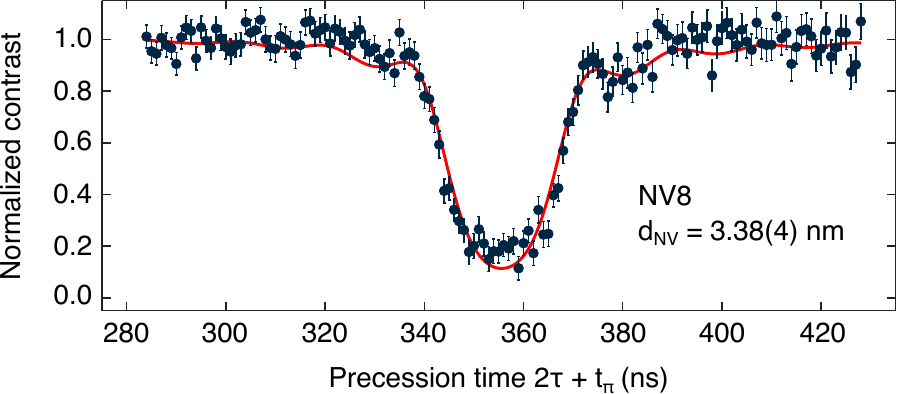}
\caption{NV-based nuclear magnetic resonance measurement with an XY8-k sensing sequence. The signal plotted here is used for measuring the depth of NV8, with immersion oil covering the diamond surface. For the plotted measurement we apply an external bias field of 330.5 Gauss aligned to the NV symmetry axis and use $k = 5$ (40 $\pi$ pulses). We supersample beyond the 2-ns timing resolution of our hardware by use of a quantum interpolation technique developed in \cite{Ajoy2017}. The solid red curve is a fit to the model developed in \cite{Pham2016} with $2\tau$ spacing between $\pi$ pulses and $t_{\pi}$ $\pi$ pulse duration, and we fit with a finite nuclear spin dephasing time $T^*_{2n}$ (here $T^*_{2n} = 26(4) ~\mu$s).}
\label{N3depth}
\end{figure}

We use NV-based nuclear magnetic resonance (NMR) to measure the increase in number of surface $^1$H in Fig.~2(a) of the main text and also to measure NV depths. In Fig.~\ref{N3depth} we plot an example signal of an NMR signal with an XY8-k sensing sequence \cite{Pham2016,Loretz2015}. The solid red curve is a fit to the model developed in \cite{Pham2016}, which yields a root-mean-square magnetic field $B_\text{RMS}$ produced by $^1$H nuclear spins (see \cite{Pham2016}). Figure~2(a) of the main text plots the fitted $B_\text{RMS}$ measured on NV1 without oil on the surface. For NV depth measurements (as presented in Fig.~\ref{N3depth} for NV8) we cover the diamond surface with Olympus Immersion Oil Type F, and we relate NV depth $d_{\text{NV}}$ to the fitted $B_\text{RMS}$ with the model from \cite{Pham2016}, assuming a $^1$H nuclear spin density of $60 / \text{nm}^3$. 

We ensure that our observed $^1$H signal comes from $^1$H and not from the 4$^{\text{th}}$ harmonic of $^{13}$C by verifying the absence of the 2$^{\text{nd}}$ harmonic of $^{13}$C \cite{Loretz2015}; we do not observe $^{13}$C signals, as expected since our NVs are hosted by an isotopically purified $^{12}$C diamond layer (see SI Note 1). We further note that we observe seemingly anomalous signals under external bias fields misaligned with the NV axis. These signals are substantially narrower in the frequency domain than would be expected for the NV filter function and do not observably shift with changes in magnetic field. We attribute these seemingly anomalous signals to harmonics produced by coherent coupling between the NV electronic spin and the host $^{14}$N nuclear spin, as demonstrated and explained in \cite{Loretz2015}, and we note that these signals disappear under external bias fields well aligned to the NV symmetry axis.

\section{Supplementary Note 7: Electron trap properties}

\begin{figure}
\includegraphics[width=125mm]{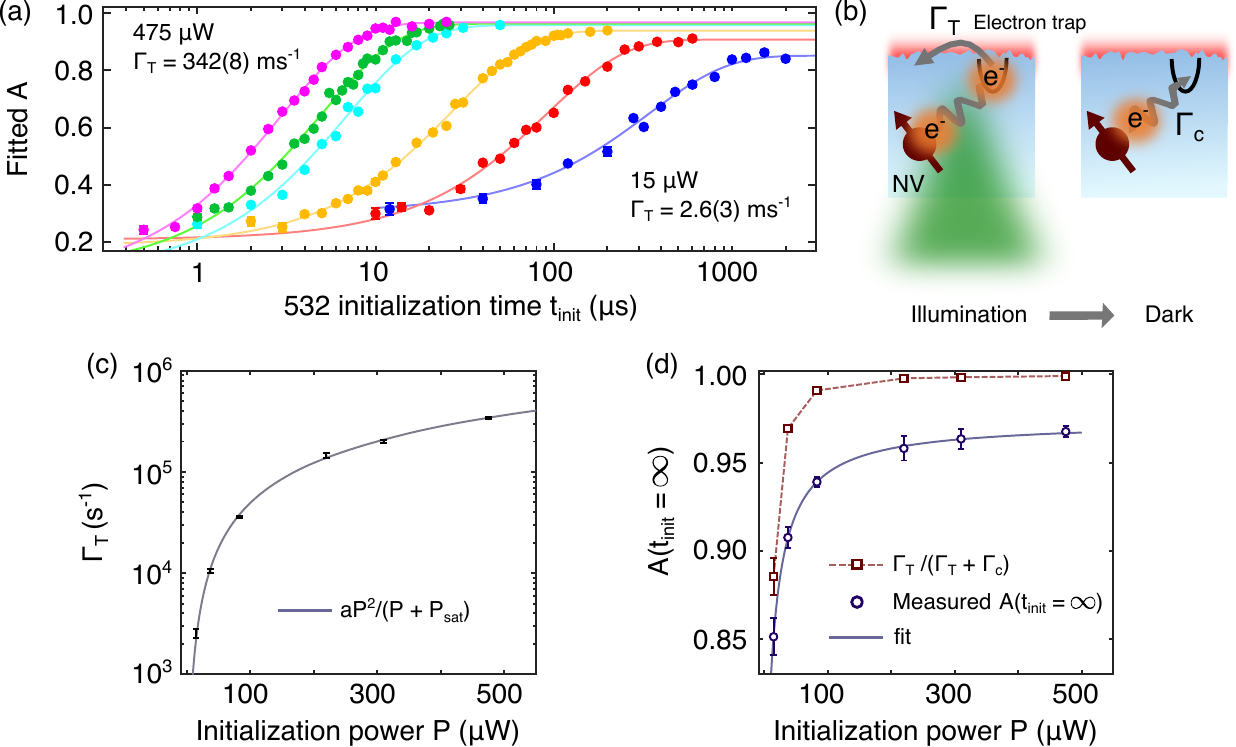}
\caption{Power-dependence of trap ionization. (a) Fitted values of NV$^-$ decay amplitude $A$ [see Eq.~(\ref{fitmodelS})] for varied initialization times at 6 laser powers (475 $\mu$W, 310 $\mu$W, 225 $\mu$W, 83 $\mu$W, 37 $\mu$W, and 15 $\mu$W from left to right). The solid curves are fits to $A(t_{\text{init}}) = - B \exp\left(-\Gamma_{\text{fit}} t_{\text{init}}\right) + A(t_{\text{init}} = \infty) $, where rate $\Gamma_\text{fit}$, saturation value $A(t_{\text{init}} = \infty)$, and amplitude $B$ are all free fit parameters. Because we find that $\Gamma_T$ can vary in time over day-to-month timescales, to minimize systematic error we measure the $A(t_{\text{init}})$ values in a quasirandomized order and measure the curves in the quasirandom order 225 $\mu$W, 83 $\mu$W, 37 $\mu$W, 310 $\mu$W, 475 $\mu$W, 15 $\mu$W. (b) Model for charge decay. (c) Trap ionization rate $\Gamma_T$ vs. 532-nm laser power. $\Gamma_T \equiv \Gamma_{\text{fit}} - \Gamma_c$, where $\Gamma_c = 332(1)$ s$^{-1}$. Solid curve is a fit to $a P^2 / (P + P_{\text{sat}})$. (d) $A(t_{\text{init}} = \infty)$ vs. 532-nm laser power. $\Gamma_T / (\Gamma_T + \Gamma_c)$ qualitatively reproduces the observed saturation behavior, but $A(t_{\text{init}} = \infty)$ saturates slower with power and to a value less than 1. The solid curve is a fit to $(b P + c) / (P + P_{\text{sat}})$ where $P$ is laser power, $b = 0.973(2)$, $c = 8(1) ~\mu$W, and $P_{\text{sat}} = 12(2)~\mu$W. We empirically find that this functional form fits the data well, which suggests that there may be other effects not fully encompassed by our model.}
\label{GammaT}
\end{figure}

In this section we present supplementary discussion of NV$^-$ decay in the dark and the observed electron trap properties. In Fig.~\ref{GammaT}(a) we plot the fitted value of $A$ from Eq.~(\ref{fitmodelS}) as a function of different initialization time and power (same data as Fig.~4(b) of main text). In our model [Fig.~\ref{GammaT}(b)] $A$ is the probability the trap is empty. To mathematically model trap ionization under illumination we consider $A(t_{\text{init}} = 0) = 0$, constant trap photoionization at rate $\Gamma_T$, and constant trap repump at rate $\Gamma_c$. These considerations yield 

\begin{equation}
A(t_{\text{init}}) = \frac{\Gamma_T}{\Gamma_T + \Gamma_c} \left(1 - e^{-\left(\Gamma_T + \Gamma_c\right) t_{\text{init}}}\right) 
\label{A_t_init}
\end{equation}
This analysis indicates that with our model the exponential saturation in Fig.~\ref{GammaT}(a) is actually at a rate $\Gamma_T + \Gamma_c$.

In Fig.~\ref{GammaT}(c) we plot the trap ionization rate $\Gamma_T$ versus laser power. We find $\Gamma_T$ is fit well by a saturation model of $a P^2 / (P + P_{\text{sat}})$, where $a$ = 814(43) s$^{-1} / \mu\text{W}$, $P$ is laser power, and the saturation power $P_{\text{sat}} = 65(23)~ \mu$W. This power-dependence is consistent with trap ionization by a two-photon transition through an orbital excited state. This two-photon behavior is sensible physically because NV$^-$ requires the energy of two 532-nm photons to photoionize \cite{Aslam2013} and we expect the trap is lower in energy than NV$^-$ in order for tunneling to be energetically favorable.

We plot the fitted value of $A(t_{\text{init}} = \infty)$ as a function of laser power in Fig.~\ref{GammaT}(d) and observe that $A(t_{\text{init}} = \infty)$ increases with power. From Eq.~(\ref{A_t_init}) we find $A(t_{\text{init}} = \infty) = \Gamma_T / (\Gamma_T + \Gamma_c)$, which qualitatively reproduces the observed saturation behavior in Fig.~\ref{GammaT}(d). However, $A(t_{\text{init}} = \infty)$ appears to saturate slower with laser power and to a value less than 1. To improve our model under illumination we consider the two NV charge states in addition to the two trap states and allow $A(t_{\text{init}} = 0) > 0$. We solve the 4-level system numerically with $A(t_{\text{init}} = 0) \approx 0.2$ and find no qualitative or quantitative difference in the behavior of $A(t_{\text{init}} = \infty)$ compared to $\Gamma_T / (\Gamma_T + \Gamma_c)$, except that $\Gamma_c \rightarrow 0.7 ~\Gamma_c$ because the NV is only in NV$^-$ 70\% of the time during the green initialization.

Most notably, no mechanism in our present model [Fig.~\ref{GammaT}(b)] prevents $A$ from reaching 1 at infinite laser power. Regardless, we observe $A$ to saturate below 1, and from the 5 NVs in Fig.~3(b) of the main text, $A$ = 0.970(1), 0.974(2), 0.972(1), 0.966(5), 0.89(1) from left to right. The similar, non-unity values of $A$ in the first 4 NVs could correspond to trap photorecombination at rate $0.03~ \Gamma_T$, preventing the trap from fully emptying under intense illumination. Instead, the non-unity $A$ could correspond to trap-NV$^0$ tunneling at rate $0.03 ~\Gamma_c$, so even if the trap is fully emptied under illumination the NV$^-$ population will never decay fully to NV$^0$. This trap-NV$^0$ tunneling at rate $0.03 ~\Gamma_c$ would represent a thermal Boltzmann factor with the trap 0.1 eV lower in energy than the NV. Performing these measurements as a function of temperature could possibly be used to probe the energy separation between NV and trap.

Physical trap candidates where tunneling could be energetically favorable include vacancy-related complexes, specifically divacancy \cite{Dhomkar2018,Deak2014} and surface sp$^2$ defects \cite{Stacey2018}. Tunneling to a nearby nitrogen substitutional defect is energetically unfavorable \cite{Farrer1969}. We do not observe a strong correlation of $\Gamma_c$ with NV depth (see Table~\ref{depthTable}). However, with a low trap density that would be consistent with the single trap behavior we observe in this work, we do not expect a strong correlation of $\Gamma_c$ with NV depth even if all traps are on the surface. Nevertheless, we do observe variations in $\Gamma_c$ and $\Gamma_T$ over day-to-month timescales or after surface treatments, indicating that the trap properties and the NV-trap tunneling mechanism depend on the surface.

\begin{table}
\centering
\begin{tabular}{|l|c|l|l|}
 \hline
 NV & Measured depth & Observed charge conversion in the dark & Location in main text \\
 \hline
 NV8   & 3.5(3) nm   & No observed charge conversion for 500 ms dark time & NA\\
 NV9 &   4.5(3) nm  & No observed charge conversion for 500 ms dark time & NA\\
 NV10 & 8.9(5) nm & No observed charge conversion for 500 ms dark time &NA \\
 NV11  &14.4(6) nm & No observed charge conversion for 500 ms dark time & NA\\
  NV2  & $>$ 4.3 nm & No observed charge conversion for 500 ms dark time & Figs.~1(d), 1(e), 1(f), and Fig.~2(c)\\
 NV1 & 3.3(3) nm & NV$^{-}$ ionization in the dark with $\Gamma_c = 0.5(2)$ s$^{-1}$ & Fig.~1(c) and Fig.~2(a) \\
 NV7 & 10.1(5) nm & NV$^{-}$ ionization in the dark with $\Gamma_c = 2.41(8)$ s$^{-1}$ & Fig.~3\\
 NV5 & $>$ 7.6 nm  & NV$^{-}$ ionization in the dark with $\Gamma_c = 332(1)$ s$^{-1}$ & Figs. 3-5\\
  NV3 & $>$ 3.0 nm  & NV$^{-}$ ionization in the dark with $\Gamma_c = 10730(30)$ s$^{-1}$ &Fig. 3\\
 \hline
\end{tabular}
\caption{NV depth does not exhibit strong correlation with charge conversion in the dark. NV depth is measured experimentally with proton NMR, as explained in Fig.~\ref{N3depth}, with depth uncertainty estimated by the variance in the fitted depth for multiple NMR measurements with different numbers of $\pi$ pulses. Coherence times and/or Rabi contrasts of NV2, NV3, and NV5 were too low to measure depth quantitatively, and so we set a lower bound by considering the maximum possible signal as the magnitude of the measured error bars in the NMR signal.  The NVs tabulated here are not meant to accurately reflect the distributions in depth or charge conversion in the dark.}
\label{depthTable}
\end{table}

\end{document}